**Giant spin Hall effect with multi-directional spin components in Ni$_4$W**

*Yifei Yang, Seungjun Lee, Yu-Chia Chen, Qi Jia, Duarte Sousa, Michael Odlyzko, Javier Garcia-Barriocanal, Guichuan Yu, Greg Haugstad, Yihong Fan, Yu-Han Huang, Deyuan Lyu, Zach Cresswell, Tony Low\*, Jian-Ping Wang\**

Yifei Yang, Seungjun Lee, Yu-Chia Chen, Qi Jia, Duarte Sousa, Yihong Fan1, Yu-Han Huang, Deyuan Lyu, Tony Low, Jian-Ping Wang
Department of Electrical and Computer Engineering, University of Minnesota, Minneapolis, MN 55455, USA
E-mail: jpwang@umn.edu; tlow@umn.edu

Michael Odlyzko, Javier Garcia-Barriocanal, Guichuan Yu, Greg Haugstad
Characterization Facility, University of Minnesota, Minneapolis, MN 55455, USA

Zach Cresswell, Jian-Ping Wang
Department of Chemical Engineering and Materials Science, University of Minnesota, Minneapolis, MN 55455,USA

**Spin-orbit torque (SOT) can be used to efficiently manipulate the magnetic state of magnetic materials, which is an essential element for memory and logic applications. Due to symmetry constraints, only in-plane spins can be injected into the ferromagnet from the underlying SOT layer for conventional SOT materials such as heavy metals and topological materials. Through the use of materials with low symmetries, or other symmetry breaking approaches, unconventional spin currents with out-of-plane polarization has been demonstrated and enabled field-free deterministic switching of perpendicular magnetization. Despite this progress, the SOT efficiency of these materials has typically remained low. Here, we report a giant SOT efficiency of 0.85 in sputtered Ni$_4$W/CoFeB heterostructure at room temperature, as evaluated by second harmonic Hall measurements. In addition, due to the low crystal symmetry of Ni$_4$W, unconventional out-of-plane and Dresselhaus-like spin components were observed.**

Macro-spin simulation suggests our spin Hall tensor to provide about an order of magnitude improvement in the magnetization switching efficiency, thus broadening the path towards energy efficient spintronic devices using low-symmetry materials.

1. Introduction

Heavy metals[1] and topological materials[2] are common SOT material candidates due to their large spin-orbit coupling (SOC). However, these materials often possess high crystal symmetry, limiting them to the so-called conventional SOT transduction, where the spin polarization (herein denoted to be along Y) is mutually orthogonal to the charge (X) and spin flow (Z) directions (**Figure 1**b). For SOT devices featuring perpendicular magnetic anisotropy (PMA), an external magnetic field is required to break the symmetry when using these high-symmetry materials as the spin-source layer[1b, 1e]. It is now well recognized that efficient switching of PMA magnet requires a finite amount of out-of-plane Z-spin, accompanying the conventional Y-spin[3]. This is because Z-spins intrinsically break the symmetry of the system and facilitate field-free switching[4]. Recently, several materials have been identified for the generation of unconventional spins with spin polarization along X and/or Z directions, including materials with low crystal symmetry[5] or magnetic symmetry[6]. Despite these advancements, the SOT efficiency of these materials remains much to be desired, limiting their competitiveness in practical applications.

The existence of unconventional spins is determined by the symmetry of the spin-source material[7]. In the framework of the linear response theory, spin Hall conductivity (SHC) can be written in the tensorial form as $\sigma_{ij}^k$ where *i*, *j* and *k* are spin polarization, spin current flow and charge current flow directions in the crystal coordinate system, respectively. Reducing the symmetry of the spin-source material will generally result in more non-zero SHC components. Therefore, careful consideration of the symmetry of the spin-source layer is crucial in the quest for materials exhibiting unconventional SHC (USHC). In this study, we identify the space group I4/m as a crystal family allowing for USHC ($\sigma_{xz}^x$) based on theory[7] and we search the material database[8] for potential candidates with stable phases within the I4/m space group. Employing density functional theory (DFT) calculations, $Ni_4W$ is identified as the most promising candidate because it is the ground state for the Ni-W binary intermetallic system[8-9] and it exhibits large theoretical SOT efficiency. We verify the existence of USHC for both $Ni_4W(100)$ and $Ni_4W(211)$ using first-principles calculations. We choose to focus our experimental efforts on the (211) orientation due to its better SOT efficiency, surpassing that of the (100) orientation. In fact, theoretical calculations confirm that Ni$_4$W (211) is about the most optimal crystal orientation for USHC. Additionally, its hexagonal-like lattice structure makes it easier to grow experimentally.

Theoretically predicted multi-directional spin components are verified experimentally. We grow epitaxial Ni$_4$W (211) films on Al$_2$O$_3$ (sapphire) (0001) substrates using magnetron sputtering. SOT from spin polarizations along X and Z directions is observed in Ni$_4$W/CoFeB heterostructures at room temperature using second harmonic Hall measurements. The SHC from Z-spin is $1.47 \times 10^4$ $\hbar/2e$ $(\Omega m)^{-1}$, which is comparable to other unconventional SOT materials[5a, 5c, 5d, 5f, 6a, 6e]. A giant conventional SOT efficiency ($\theta_{DL}^Y$) as large as 0.85 is also observed, which can greatly improve the magnetization switching efficiency[3b]. Macro-spin simulations suggest that the multi-directional spins of Ni$_4$W collectively enhance the switching efficiency, rendering it superior to that of other unconventional SOT materials.

## 2. Results and Discussion

The crystal structure of Ni$_4$W is shown in Figure 1a, left subfigure, which is tetragonal with space group I4/m and lattice constants of a=b=5.73 Å and c=3.55 Å. Due to its low crystalline symmetry, Ni$_4$W can have multiple non-zero USHC components, such as $\sigma_{xz}^x$, and $\sigma_{yz}^y$, which can serve as Z-spin components for (100) and (010) orientations respectively[7b]. However, the difficulty of finding suitable substrates for (100) or (010)-oriented Ni$_4$W makes it unfeasible to utilize these components. Instead, (211)-oriented Ni$_4$W lacks mirror or rotational symmetry relative to the surface (Figure 1a, right subfigure), meeting the requirements for Z-spin generation[10], and can also be experimentally grown on hexagonal substrates. Symmetry analysis and first-principles calculations verified that unconventional out-of-plane (Z), Dresselhaus-like (X), and conventional (Y) spins can be generated in Ni$_4$W (211) (Supplementary sections 1 and 2), and that this orientation is also most optimal (Figure S2 in Supplementary section 2). Figure 1b illustrates the spin current generation in Ni$_4$W/CoFeB heterostructure.

We performed first-principles calculations to study the band structure and SHC of Ni$_4$W. Figure 1c shows the electronic structure of Ni$_4$W including SOC. In the absence of SOC, it contains multiple Dirac nodal lines due to the time reversal, inversion, and $M_{xy}$ mirror symmetries[11]. Then, with SOC turned-on, strong SOC due to W opens a gap along the Dirac nodal lines, resulting in spin-Berry curvature hotspots as represented in Figure 1c[12]. Multiple spin-Berry curvature hotspots give rise to a giant SHC in Ni$_4$W. The biggest SHC tensor component of (001)-oriented Ni$_4$W (Ni$_4$W (001)) is calculated to be $\sigma_{xy}^z = 1.36 \times 10^5$ $\hbar/2e$ $(\Omega m)^{-1}$ at the Fermi level. Although this value is around one-third of that of Pt ($\sigma_{xy}^z = 4.42 \times 10^5$ ℏ/2e (Ωm)$^{-1}$), the electrical conductivity of Ni$_4$W is one order smaller than that of Pt, implying a higher spin Hall angle in Ni$_4$W. The electrical conductivity of Ni$_4$W (001) is calculated to be $\sigma_{yy} = 7.1 \times 10^5$ (S/m), resulting in a large spin Hall angle of $\theta^Y$=0.192. This value is three times larger than that of Pt evaluated through the same method.

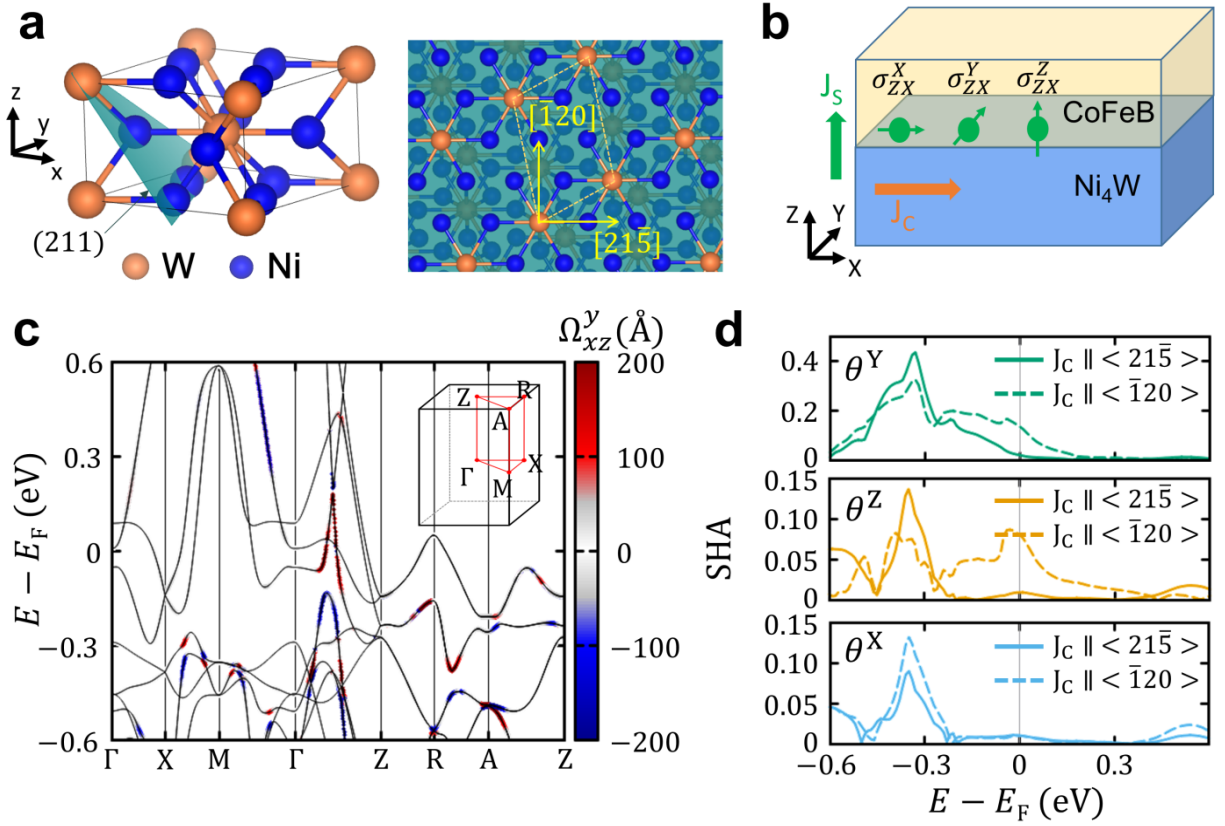

**Figure 1.** Structure of Ni$_4$W and DFT calculations. a) Crystal structure of Ni$_4$W with a tetragonal lattice. One of the (211) planes is shown. In the top view of Ni$_4$W (211), a unit cell with an oblique shape is presented, lacking rotational or mirror symmetry. Two orthogonal unit vectors are also labelled. b) Schematic of Ni$_4$W(211)/CoFeB heterostructure and multi-directional spin components. c) Electronic structure of Ni$_4$W with SOC. Inset shows its 1$^{st}$ Brillouin zone and the spin Berry curvature component of $\Omega_{xz}^y$ is represented as color coordinates. d) Spin Hall angles of Ni$_4$W (211). Green, yellow, and blue colors indicate $\theta^Y$, $\theta^Z$, and $\theta^X$, respectively, and solid and dashed lines represent two orthogonal current directions.

Through standard rotation operation on the conductivity tensors of Ni$_4$W (001), we calculated the magnitude of spin Hall angle of Ni$_4$W (211) as shown in Figure 1d. Due to its anisotropic in-plane crystal orientation, the magnitude of SHC tensor varies according to the current direction. When $J_C \parallel <\bar{1}20>$, we observed sizable $\theta^Y$=0.134 and $\theta^Z = 0.08$ at Fermi level, suggesting that Ni$_4$W (211) can be a promising SOT material for energy-efficient spintronic devices. Although the calculated SHA are relatively small when $J_C \parallel <21\bar{5}>$, we still can find non-zero unconventional SHA in this orientation. Notably, we found that the calculated SHA strongly depends on the position of Fermi level. The $\theta^Y, \theta^Z,$ and $\theta^X$ are simultaneously maximized near $E = E_F - 0.33$ eV, and can approach 0.435, 0.136, and 0.132, respectively. This suggests that the obtained SOT can be further optimized through hole doping, opening a feasible way to optimize conventional and unconventional SOT of Ni$_4$W.

Epitaxial Ni$_4$W thin films were grown on Al$_2$O$_3$ (sapphire) (0001) substrates using magnetron sputtering at 350 °C. A thin W seed layer with thickness of approximately 2 nm was used between the substrate and Ni$_4$W layer. For charge-to-spin studies, Ni$_4$W/CoFeB samples were prepared, wherein a 5 nm CoFeB layer was deposited on top of Ni$_4$W after cooling down. Ta or SiO$_2$ were deposited as capping layers to prevent oxidation of the metallic layers. A reference sample of Al$_2$O$_3$/W (3 nm)/CoFeB (5 nm)/cap was also prepared. Ni$_4$W composition was confirmed by Rutherford backscattering spectrometry (RBS) (Supplementary section 3). Vibrating-sample magnetometry (VSM) of Ni$_4$W samples shows the non-magnetic nature of Ni$_4$W (Supplementary section 4). Surface roughness of Ni$_4$W was measured as 0.33 nm for the Al$_2$O$_3$/W (2 nm)/Ni$_4$W (10 nm) sample (Supplementary section 5) by atomic force microscopy (AFM), confirming suitability for Ni$_4$W/CoFeB integration and interfacial spin transport.

A thorough X-ray diffraction (XRD) study has been done to verify the phase and orientation of Ni$_4$W phase as well as to assess the crystalline quality of the films. As shown in **Figure 2**a, the specular wide angle XRD presents an intense Ni$_4$W (211) peak at 2θ = 43.45° besides the substrate peak, which confirms the intended phase and orientation of the Ni-W film in Al$_2$O$_3$ (0001)/W (2 nm)/Ni$_4$W (30 nm)/CoFeB (5 nm)/cap. At lower angles, a broad peak corresponding to the thin W seed layer with (110) orientation, along with its corresponding finite thickness oscillation, can also be observed. The inset in Figure 2a shows the rocking curve (omega scan) of the Ni$_4$W (211) peak. The full width at half-maximum (FWHM) value is below 0.1°, which confirms the high crystalline quality of the samples showing a small degree of crystal mosaicity. Rocking curves of samples of different Ni$_4$W thicknesses (Supplementary section 6) show an increasing spread of the crystal plane orientations for increasing Ni$_4$W thickness. Nevertheless, the FWHM values of these samples are all below 0.2°, indicating highly coherent Ni$_4$W growth. The KL projection of the 3D reciprocal space map of the same sample is shown in Figure 2c. The map is referred to the crystallographic reference of the substrate and shows the sample's specular rod of scattering for the same region as the scan shown in Figure 2a. We can observe the $Al_2O_3$ (0006) substrate peak with the broad W (110) peak centered around L=5.9 r.l.u., as well as the intense $Ni_4W$ (211) peak. The shape of the peaks projected in both this KL projection and the HL projection of the data (see supplementary section 7) is remarkably narrow, with peak widths along the K (Figure 2c) and H (Figure S7a) directions as short as the one observed for the substrate. This evidences a highly ordered epitaxial growth of both layers with a large lateral structural coherence and small degree of disorder.

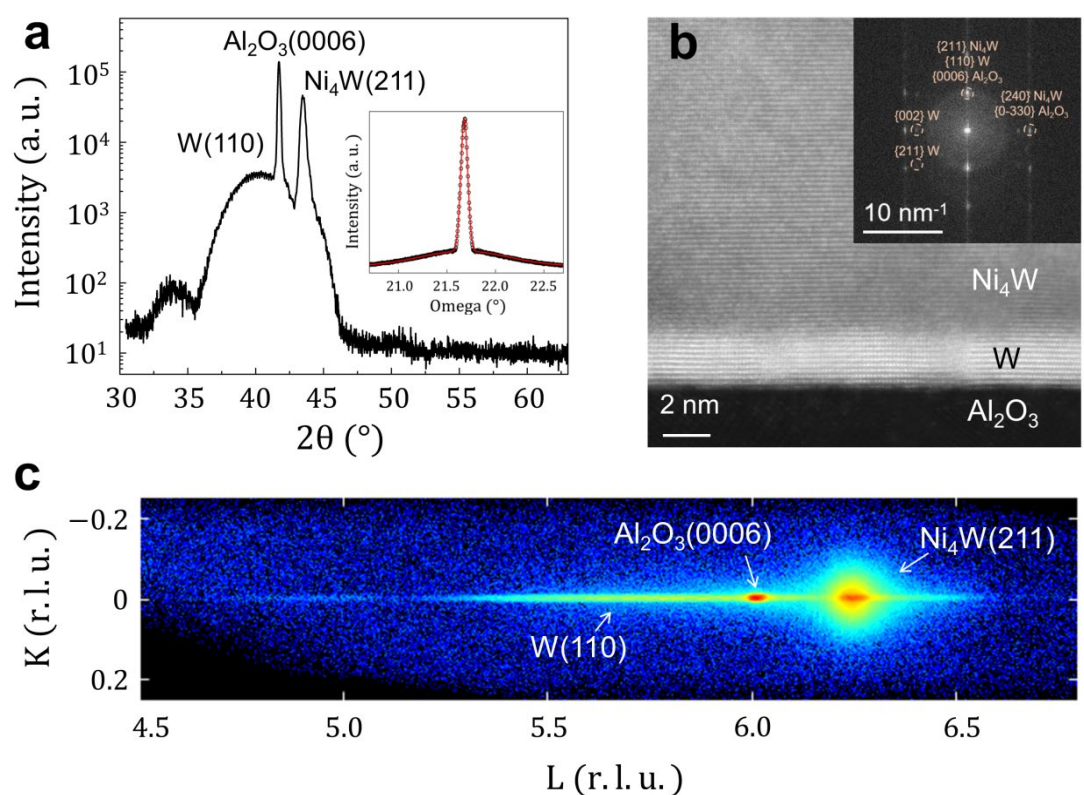

**Figure 2.** Structural characterization of $Al_2O_3/W/Ni_4W/CoFeB$. a) XRD pattern of $Al_2O_3$ (0001)/W (2 nm)/$Ni_4W$ (30 nm)/CoFeB (5 nm)/cap. Inset: XRD rocking curve of $Ni_4W$ (211) peak with FWHM of 0.084°. b) HAADF-STEM image obtained from the $<11\bar{2}0>$ zone axis of sapphire. Inset: corresponding fast-Fourier transformation (FFT). c) XRD reciprocal space mapping of $Al_2O_3$ (0001)/W (2 nm)/$Ni_4W$ (30 nm)/CoFeB (5 nm)/cap in units of sapphire.

To further study the microstructure of $Al_2O_3$(0001)/W (2 nm)/$Ni_4W$ (30 nm)/CoFeB (5 nm)/cap, we also performed scanning transmission electron microscopy (STEM). Low-magnification STEM imaging (Supplementary section 8 Figure S8a) of the film stack cross-section confirms the growth of smooth layers with flat interfaces. Energy-dispersive X-ray spectroscopy (EDS) (Supplementary section 8 Figure S8b) confirms the growth of stoichiometric Ni$_4$W. High-magnification STEM imaging reveals in-plane texturing relationships between the Ni$_4$W film, W seed layer and the sapphire substrate, further confirming the epitaxial growth of the W and Ni$_4$W films (Supplementary section 8 Figure S9). High-angle annular dark-field (HAADF) imaging (Figure 2b) directly shows the high-quality structure of the sapphire/W and W/Ni$_4$W interfaces. Highly ordered W and Ni$_4$W crystal lattices are clearly observed, corresponding to a highly anisotropic dot profile in the image FFT (inset of Figure 2b). Complementing standard cross-section imaging, plan-view STEM imaging directly shows the full in-plane structure of both the W and Ni$_4$W layers (Supplementary section 8 Figure S10).

To study the charge-to-spin conversion of Ni$_4$W, we carried out second harmonic Hall (SHH) measurements at room temperature. The Ni$_4$W/CoFeB samples and the reference sample (W/CoFeB) were patterned into Hall bar devices 110 μm long and 10 μm wide using optical lithography. The resistivity values at room temperature of Ni$_4$W and CoFeB were measured by four-probe transport measurements to be 150 and 227 μΩ cm, on Al$_2$O$_3$/Ni$_4$W(30 nm)/SiO$_2$ and Al$_2$O$_3$/CoFeB(5 nm)/SiO$_2$, respectively. The SHH setup is illustrated in **Figure 3**a, where an alternating current is applied in the longitudinal direction of the Hall bar and the Hall voltage is measured in the transverse direction by a lock-in amplifier (LIA). An external magnetic field in the plane of the device is applied at an angle $\varphi$ relative to the current direction. Spin currents are generated from the spin-source layer and the resulting SOT oscillates the magnetization of the CoFeB, the magnitude of which can be extracted from the Hall voltage signals[5c, 6e, 13]:

$$V_\omega = V_P \sin 2\varphi \tag{1}$$

$$V_{2\omega} = V_{DL}^Y \cos\varphi + V_{DL}^X \sin\varphi + V_{DL}^Z \cos 2\varphi + V_{FL}^Y \cos\varphi \cos 2\varphi + V_{FL}^X \sin\varphi \cos 2\varphi + V_{PNE} \sin 2\varphi + V_{FL}^Z \tag{2}$$

where $V_P$ is the planar Hall effect (PHE) voltage; $V_{DL}^X$, $V_{DL}^Y$, $V_{DL}^Z$ are components from damping-like torques generated by X, Y and Z spins, respectively; $V_{FL}^X$, $V_{FL}^Y$, $V_{FL}^Z$ are the field-like counterparts for those spins; $V_{PNE}$ is the planar Nernst effect (PNE) voltage. Figure S11b shows the first harmonic and Figure 3b shows the second harmonic signal of Ni$_4$W (5 nm)/CoFeB (5 nm) sample under the magnetic field of 4000 Oe, at room temperature with the current applied along <$\bar{1}$20> direction of Ni$_4$W. Non-zero $V_{DL}^X$ and $V_{DL}^Z$ values evidence the existence of X- and Z-spins.

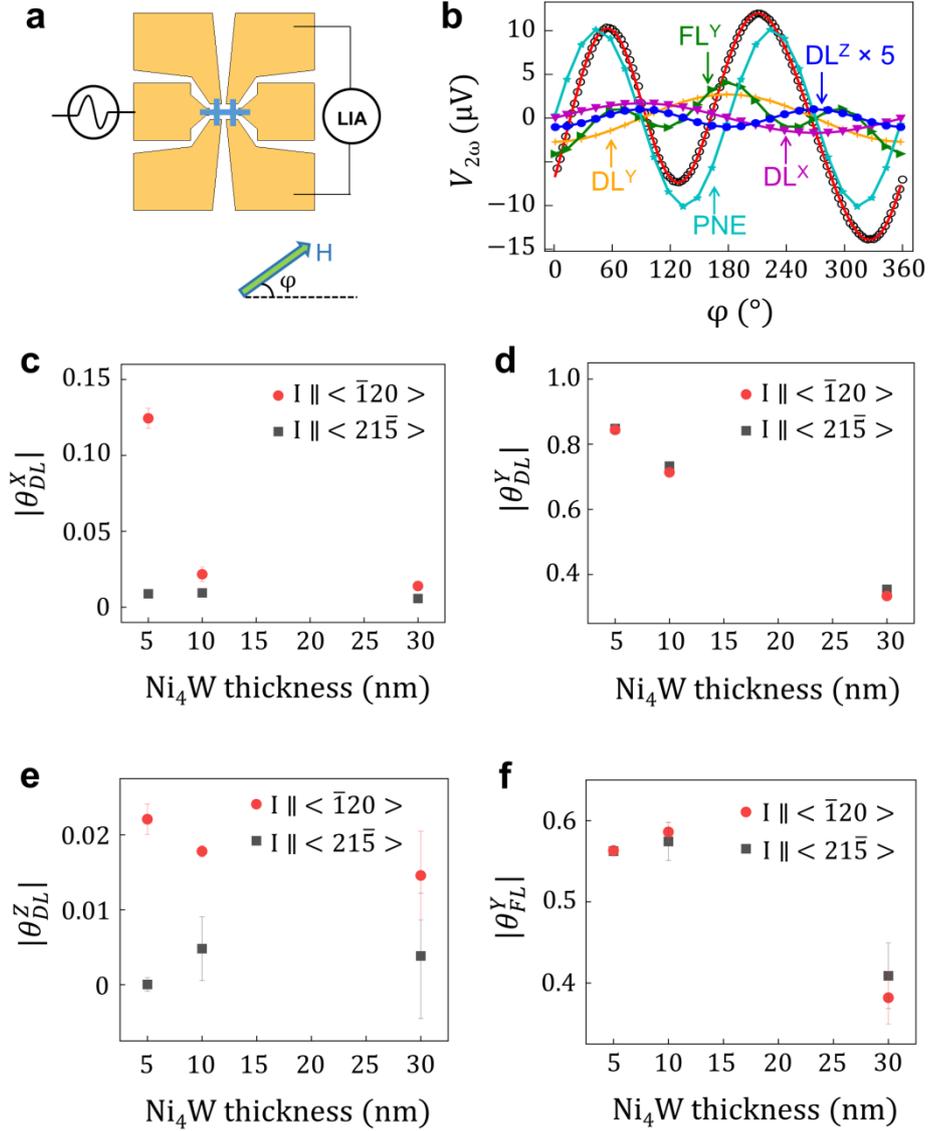

**Figure 3.** Charge-to-spin conversion measurements of Ni$_4$W/CoFeB heterostructures. a) Schematic of harmonic Hall measurement setup. b) Second harmonic voltage $V_{2\omega}$ for the Ni$_4$W (5 nm)/CoFeB (5 nm) sample, under magnetic field of 4000 Oe at room temperature, with current applied along <$\bar{1}20$> axis of Ni$_4$W. Fitting results of second harmonic voltage show that X- and Z-spins exist in the sample. c-f) SOT efficiencies of Al$_2$O$_3$(0001)/W (2 nm)/Ni$_4$W (t nm)/CoFeB (5 nm)/cap samples with different Ni$_4$W thicknesses from damping-like torque of X-spin (c), Y-spin (d), Z-spin (e) and field-like torque of Y-spin (f).

In order to evaluate SOT efficiencies for each the three spin components, we conducted SHH measurements with different magnetic fields in the range of 3,000 Oe to 10,000 Oe. Components from X-, Y- and Z-spins can be clearly seen to vary linearly with the inverse of field for Ni$_4$W (5 nm)/CoFeB (5 nm) (Supplementary section 9 Figure S12). The anomalous Hall resistance of the Ni$_4$W/CoFeB heterostructure was measured to be 0.57 Ω and the anisotropy field was 10,000 Oe, both of which were obtained from Hall resistance measurements with a perpendicular field sweep. Based on analysis of the field-dependence of voltage components in equation (2) (details in Supplementary section 9), the damping-like SOT efficiencies for X-, Y- and Z-spins ($\theta_{DL}^X, \theta_{DL}^Y, \theta_{DL}^Z$) are 0.124 ±0.007, 0.849 ±0.016, and -0.022 ±0.002, respectively. The field-like counterparts for Y- and Z-spins ($\theta_{FL}^Y, \theta_{FL}^Z$) are 0.577 ±0.008, and 0.025 ±0.003, respectively. The field-like torque of X-spin ($\theta_{FL}^X$) is considered negligible, as there is no clear linear dependence observed on the external field. Spin Hall conductivity (SHC) is calculated by $\theta/\rho_{XX}$, where $\theta$ is the SOT efficiency and $\rho_{XX}$ is charge resistivity. The damping-like SHC of Z-spin is evaluated to be $1.47 \times 10^4$ ℏ/2e (Ωm)$^{-1}$, which is comparable to other unconventional SOT materials such as WTe$_2$[5a], RuO$_2$[6e], Mn$_3$GaN[6a] and MnPd$_3$[5c]. In contrast, the W reference sample only exhibits conventional Y-spin, and the damping-like and field-like SOT efficiencies are measured to be -0.132 ± 0.003 and 0.036 ± 0.001, respectively, which are similar to the values previously reported[1d] (Supplementary section 9 Figure S13).

Experimentally determined SOT efficiencies of X- and Y-spins are greater than the predictions by DFT calculation. Physically, in addition to the intrinsic SHC, we may have additional contributions from Rashba-Edelstein effects at the W/Ni$_4$W interface that can contribute to $\theta^X$ and $\theta^Y$. The other plausible scenario is the doping effects. Figure 1d clearly shows that the SHC components for $\theta^X$ and $\theta^Y$ rapidly increase when Fermi level ($E_F$) moves down. Therefore, additional contributions from both Rashba-Edelstein effects and doping effects can increase both $\theta^X$ and $\theta^Y$, resulting in giant SOT. To investigate the role of the interfaces, we performed SHH measurements on the Ni$_4$W/CoFeB samples with Ni$_4$W thicknesses of 5 nm, 10 nm and 30 nm. The alternating current was applied either along the <21$\bar{5}$> or <$\bar{1}$20> direction of Ni$_4$W. As shown in Figure 3d and f, $\theta_{DL}^Y$ and $\theta_{FL}^Y$ decrease with increasing Ni$_4$W thickness. This implies that the large $\theta_{DL}^Y$ and $\theta_{FL}^Y$ in thinner samples could originate from the combination effect of the W seed layer and the Ni$_4$W layer, such as the Rashba effect at the W/Ni$_4$W interface. For the thickest 30 nm sample, spin currents generated from Rashba effects are dissipated and cannot enter the CoFeB layer. Therefore, we believe that the $\theta_{DL}^Y$ and $\theta_{FL}^Y$ are determined by the Ni$_4$W layer for the 30 nm sample. For unconventional torque efficiencies, we observed a distinct anisotropic behavior of $\theta_{DL}^Z$ for the two current directions for all samples, agreeing with DFT calculations. We also found a significant discrepancy in $\theta_{DL}^X$ for the 5 nm sample between the two current directions. To investigate the reason, we performed SHH for various current directions on the 5 nm sample (Supplementary section 9 Figure S14). The SOT efficiencies display a pronounced dependence on the current direction relative to the Ni$_4$W crystal axis, providing clear evidence of unconventional SOT originating from the reduced crystal symmetry of the highly coherent and sharply textured Ni$_4$W film.

We summarized the conventional ($\theta_{DL}^Y$) and unconventional ($\theta_{DL}^Z$) SHAs of state-of-the-art SOT materials in **Figure 4**a, including conventional ones PtAu[14], PtPd[15], PtCr[16] and also unconventional ones WTe$_2$[5a], Mn$_3$GaN[6a], RuO$_2$[6e], MnPd$_3$[5c], TaIrTe$_4$[5d], and PtTe$_2$/WTe$_2$[5f]. Ni$_4$W exhibits the largest $\theta_{DL}^Y$ of all; it also displays sizable $\theta_{DL}^Z$, surpassed only by TaIrTe$_4$. Previous studies[3b] have shown that optimal magnetization switching occurs with larger $\theta_{DL}^Y$, while accompanied by finite $\theta_{DL}^Z$ on the order of $\theta_{DL}^Z \sim 0.1\,\theta_{DL}^Y$. Fortuitously, this coincides with what we found here experimentally for Ni$_4$W.

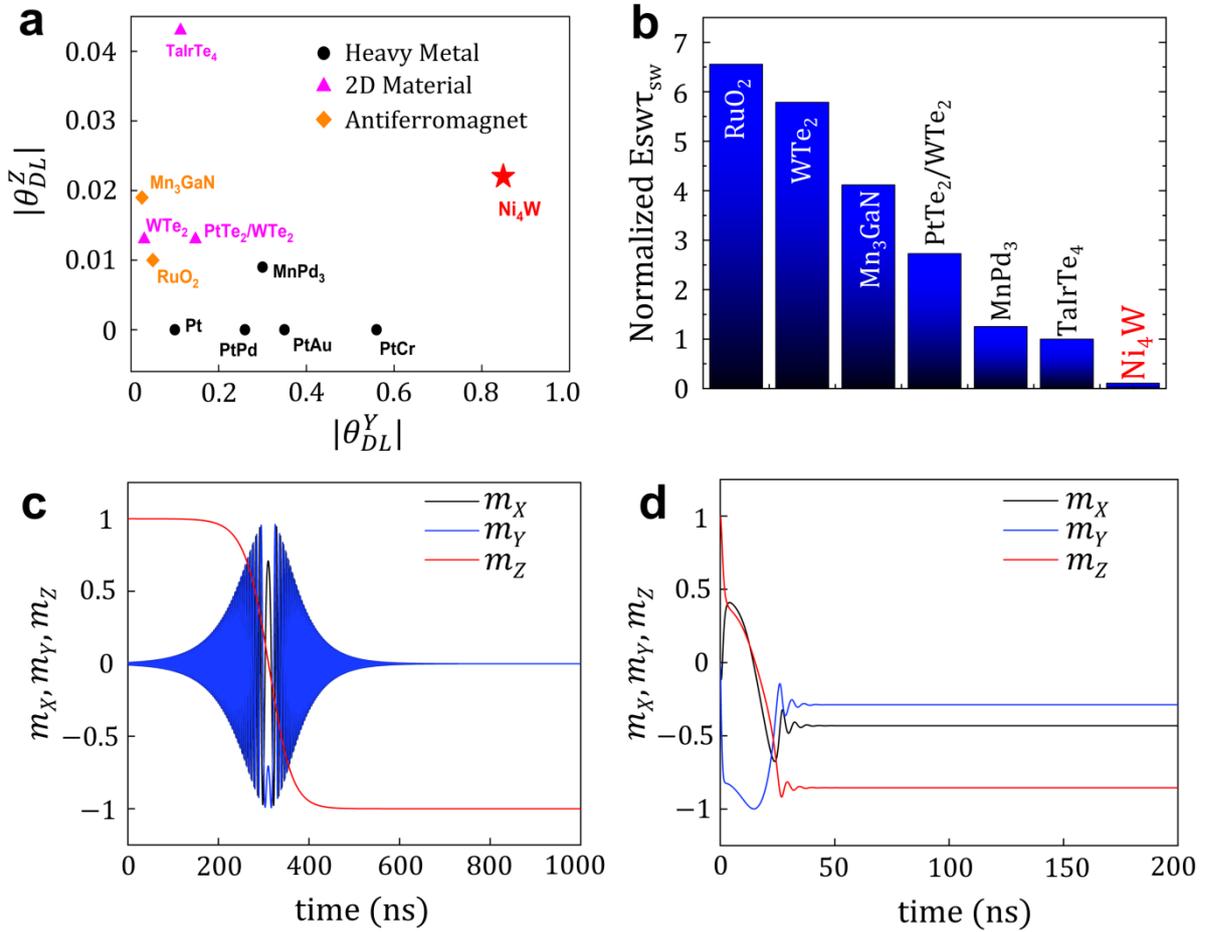

**Figure 4.** Macro-spin simulations. a) Summary of conventional and out-of-plane SHAs of Ni$_4$W compared to state-of-the-art SOT materials. b) Normalized energy delay products for perpendicular magnet switching calculated by macro-spin simulations of unconventional SOT materials. c-d) Magnetic dynamics using Ni$_4$W as SOT material with Z-spin only (c) and all X-, Y-, Z-spins (d).

To verify the performance merits of Ni$_4$W for SOT applications, we performed magnetization dynamics simulations based on the Landau-Lifshitz-Gilbert (LLG) equation[3b]. Within the macro-spin approximation, we access the field-free switching efficiency of a perpendicular magnet subjected to conventional and unconventional SOTs. Here, we quantify the switching efficiency through the numerically calculated energy delay product ($E_{sw}\tau_{sw}$) of various unconventional SOT materials relative to that of TaIrTe$_4$, a large Z-spin material. The switching time $\tau_{sw}$ is obtained numerically by registering the time spent for full magnetization reversal according to the criteria $m_Z(\tau_{sw}) < -0.1$ and $(dm_Z/dt)_{t=\tau_{sw}} < 0.01$, where $m_Z$ refers to the out-of-plane component of the magnetization that corresponds to the direction of the magnetic anisotropy, and the injected current density is taken to be $J = 3 \times 10^7$ Acm$^{-2}$. The SOT includes all three possible damping-like and field-like components, whose magnitudes are defined by the SHAs provided in Supplementary Table S3 (calculation details in Supplementary section 10). The main results are compiled in Figure 4b, where it is apparent that field-free switching is made considerably more efficient through the multi-directional spins induced by Ni$_4$W. Relative to TaIrTe$_4$, switching efficiency is enhanced by an order of magnitude. Furthermore, we find that X- and Y-spins also provide efficiency enhancement in the case of Ni$_4$W. In Figure 4c and 4d, we compare the time evolution of the magnetization components assuming only SOTs due to Z-spin of Ni$_4$W and the full multicomponent case. Fully deterministic magnetization reversal is greatly improved by an order of magnitude by the presence of SOTs due to the X and Y spin components. Hence, the simultaneous presence of sizable multi-directional spins induced by Ni$_4$W enables versatile and efficient field-free switching ferromagnetic thin films with PMA. Simulations on a magnet with in-plane anisotropy also indicate the high switching efficiency of Ni$_4$W compared to other conventional or unconventional SOT materials (Supplementary section 10 Figure S15).

## 3. Conclusion

In summary, we have identified Ni$_4$W as an unconventional SOT material, supported by DFT calculations. By fabricating epitaxial Ni$_4$W thin films, we have successfully demonstrated the giant spin Hall effect with multi-directional spin components using second harmonic Hall measurements. The unconventional SHC from damping-like Z-spin is evaluated to be $1.47 \times 10^4$ $\hbar/2e$ $(\Omega m)^{-1}$ at room temperature, which is comparable to other state-of-the-art unconventional materials. In addition, giant conventional damping-like and field-like efficiencies ($\theta_{DL}^Y, \theta_{FL}^Y$) are measured to be 0.85 and 0.58, respectively, enabling improved switching efficiency. By studying the thickness dependence of SOT efficiencies, we found that thinner Ni$_4$W samples exhibit larger $\theta_{DL}^Y$ and $\theta_{FL}^Y$ that probably originate from the W/Ni$_4$W interface. The unconventional $\theta_{DL}^Z$ corresponding to the intrinsic spin Hall effect of Ni$_4$W exhibits clear dependence on current direction and it demonstrates that the unconventional SOT originates from the low symmetry of Ni$_4$W. Macro-spin simulations on Ni$_4$W and other unconventional SOT materials indicate that Ni$_4$W has superior efficiency for magnetization manipulation. Our results present a promising unconventional SOT material candidate for energy-efficient spintronic devices.

## 4. Experimental Section

*Sample growth and device fabrication*: Ni$_4$W samples were prepared on single crystal Al$_2$O$_3$ (0001) substrates by magnetron sputtering with base pressure better than 5 ×10$^{-7}$ Torr. Both W seed layers and Ni$_4$W layers were deposited at a substrate temperature of 350 °C. Ni$_4$W layers were prepared by co-sputtering Ni and W targets at a deposition rate of 0.97 Å/s. CoFeB and capping layer were deposited after the sample was cooled down to room temperature. The reference sample with the stack of Al$_2$O$_3$/W(3 nm)/CoFeB(5 nm)/cap was prepared with W deposited at 350 °C and other layers at room temperature.

For resistivity and SHH measurements, thin film stacks were patterned into Hall bar microstrips with length of 110 μm and width of 10 μm by optical lithography and ion milling. Electrodes composed of Ti (10 nm)/Au (150 nm) were deposited on the microstrips by e-beam evaporation.

*XRD and STEM characterization*: XRD and reciprocal space mapping measurements were performed on a two-dimensional Bruker D8 Discover with Co-Kα (λ = 1.79 Å) radiation. Peak positions of the 2θ scan are converted those of a wavelength of Cu-Kα source (λ = 1.54 Å). Rocking curve studies were carried out using a Rigaku Smartlab with a Cu-Kα source. Cross-section and plan-view TEM samples were prepared using a FEI Helios NanoLab G4 dual beam FIB operated at accelerating voltages of 30, 5, 2, and 1 kV. STEM-EDS characterization was performed using a Thermo Fisher Talos F200X G2 TEM operating at an accelerating voltage of 200 kV and equipped with a Fischione 3000 HAADF detector, Thermo Fisher Super-X G2 EDS detector, and Thermo Fisher Velox software. High resolution STEM imaging was performed using a FEI Titan G2 60-300 TEM operating at an accelerating voltage of 300 kV and equipped with a CEOS DCOR probe corrector, Fischione 3000 HAADF detector, FEI BF detector, and Thermo Fisher TIA software.

*Second harmonic measurements*: Second harmonic measurements were performed using a Quantum Design physical property measurement system. A Keithley 6221 was used to provide the alternating current and two SR-830 lock-in amplifiers were used to detect the first and second harmonic signals.

**Supporting Information**

Supporting Information is available from the Wiley Online Library or from the author.

**Acknowledgements**

This work was supported, in part, by SMART, one of the seven centers of nCORE, a Semiconductor Research Corporation program, sponsored by the National Institute of Standards and Technology (NIST) and by the Global Research Collaboration (GRC) Logic and Memory program, sponsored by Semiconductor Research Corporation (SRC). Parts of this work were carried out in the Characterization Facility, University of Minnesota, which receives partial support from the NSF through the MRSEC (Award Number DMR-2011401) and the NNCI (Award Number ECCS-2025124) programs. Portions of this work were conducted in the Minnesota Nano Center, which is supported by the National Science Foundation through the NNCI under Award Number ECCS-2025124.

**Author Contributions**

J.P.W. initiated and coordinated the project. Y.Y. and J.P.W. identified the material and designed the experiments. S.L. and T.L. performed first-principles and spin transport calculations. Y.Y. grew the samples and carried out X-ray diffraction with Z.C. Y.C. patterned the samples into Hall bar devices. Y.Y. performed second harmonic measurements and analyzed the data with Q.J., Y.F. and Y.H. Q.J. performed VSM and AFM measurements. Q.J. and G.H. analyzed the AFM data. M.O. performed the TEM study. D.S. and T.L. carried out macro-spin simulations. J.G.B. performed reciprocal space mapping and analyzed the data with G.Y. G.H. carried out Rutherford backscattering spectrometry. Y.Y. drafted the manuscript with S.L. and D.S.. J.P.W. and T.L. coordinated the manuscript preparation. All authors discussed and commented on the manuscript. Y.Y. and S.L. contributed equally to this work.

M.-Y. Song, Y.-L. Huang, S.-J. Lin, C.-F. Pai, *ACS Applied Electronic Materials* **2022**, 4, 1099.

**Supporting Information**

**Giant spin Hall effect with multi-directional spin components in Ni$_4$W**

*Yifei Yang, Seungjun Lee, Yu-Chia Chen, Qi Jia, Duarte Sousa, Michael Odlyzko, Javier Garcia-Barriocanal, Guichuan Yu, Greg Haugstad, Yihong Fan, Yu-Han Huang, Deyuan Lyu, Zach Cresswell, Tony Low\*, Jian-Ping Wang\**

Yifei Yang, Seungjun Lee, Yu-Chia Chen, Qi Jia, Duarte Sousa，Yihong Fan1, Yu-Han Huang, Deyuan Lyu, Tony Low, Jian-Ping Wang
Department of Electrical and Computer Engineering, University of Minnesota, Minneapolis, MN 55455, USA
E-mail: jpwang@umn.edu; tlow@umn.edu

Michael Odlyzko, Javier Garcia-Barriocanal, Guichuan Yu, Greg Haugstad
Characterization Facility, University of Minnesota, Minneapolis, MN 55455, USA

Zach Cresswell, Jian-Ping Wang
Department of Chemical Engineering and Materials Science, University of Minnesota, Minneapolis, MN 55455,USA

**Contents:**

## 1. Theoretical calculations

We performed first-principles calculations based on density functional theory (DFT)[1] as implemented in the Quantum Espresso package (QE)[2]. The exchange-correlation functional was treated within the generalized gradient approximation of Perdew-Burke-Ernzerhof (PBE)[3]. The kinetic energy cutoff of electronic wavefunctions and charge density were chosen to be 90 and 720 Ry, respectively. The crystal structure of Ni$_4$W was fully relaxed using DFT calculations and the lattice constants were calculated to be a=b=5.739 Å and c=3.566 Å, which are in good agreement with our experimental values. The spin-orbit coupling was included in the electronic structure calculations. The k-grid meshes were chosen to be 8×8×14.

The intrinsic spin Hall conductivity $\sigma_{\alpha\beta}^{\gamma}$ was calculated by the Kubo-Greenwood formula as[4]

$$\sigma_{\alpha\beta}^{\gamma}(\varepsilon_\text{F}) = \frac{\hbar}{\Omega_C N_k}\sum_{\bm{k}}\sum_{n} f_{n\bm{k}} \sum_{m\neq n} \frac{2\text{Im}[\langle n\bm{k}|\hat{j}_{\alpha}^{\gamma}|m\bm{k}\rangle\langle m\bm{k}|-e\hat{v}_\beta|n\bm{k}\rangle]}{(\varepsilon_{n\bm{k}} - \varepsilon_{m\bm{k}})^2 + \eta^2},$$

where $\varepsilon_\text{F}$, $\Omega_C$ and $N_k$ indicate the Fermi energy, cell volume, and the number of k-points used for k-space sampling, $n$ and $\bm{k}$ are band index and crystal momentum, $f_{n\bm{k}}$ indicates Fermi-Dirac distribution, $\hat{j}_{\alpha}^{\gamma}$ is the spin current operator defined as $\hat{j}_{\alpha}^{\gamma} = \frac{1}{2}\{\hat{s}^{\gamma}, \hat{v}_\alpha\}$ and $\hat{s}^{\gamma}$ and $\hat{v}_{\alpha,\beta}$ are spin and velocity operators, respectively. $\eta$ is an adjustable smearing parameter. The numerical calculations were performed using the Wannier90 package[5]. For wannierization, s and d orbital projections were used for Ni and only d orbital was used for W. To obtain the converged results, we used a fine k-mesh grid of 61×61×101 for the bulk Ni$_4$W with 2×2×2 adaptive sampling method. The corresponding broadening constant of $\eta$ was chosen to be 10 meV. Topological nodal line structures were calculated numerically using the WannierTools package[6].

The electrical conductivity $\sigma_{\alpha\beta}$ was evaluated using Boltzmann transport equation in the constant relaxation time approximation[7] as

$$\sigma_{\alpha\beta}(\varepsilon_F, T) = e^2 \int_\infty^\infty d\varepsilon \left(-\frac{\partial f_{n\mathbf{k}}(T)}{\partial \varepsilon}\right) \Sigma_{\alpha\beta}(\varepsilon_F)$$

and $\Sigma_{\alpha\beta}(\varepsilon_F)$ is the transport distribution function tensor, defined as,

$$\Sigma_{\alpha\beta}(\varepsilon_F) = \frac{1}{\Omega_C N_k} \sum_{\mathbf{k}} \sum_n v_{n\mathbf{k}}^\alpha v_{n\mathbf{k}}^\beta \tau \delta(\varepsilon_F - \varepsilon_{n\mathbf{k}}),$$

where $\tau$ indicates the constant relaxation time and $v_{n\mathbf{k}}^\alpha$ is a group velocity of eigenstate of $n\mathbf{k}$ calculated as $v_{n\mathbf{k}}^\alpha = \langle n\mathbf{k}|\hat{v}_\alpha|n\mathbf{k}\rangle$.

## 2. Spin Hall conductivity for Ni$_4$W (001) and (211)

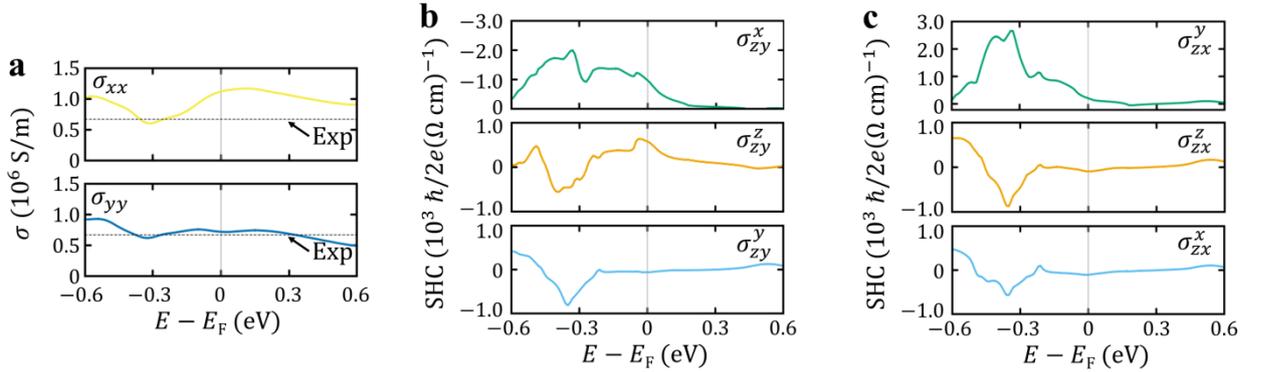

**Figure S1.** Representative components of charge conductivity (**a**) and spin Hall conductivities (**b**,**c**) of Ni$_4$W (211). In **a**, the constant relaxation time of $\tau = 5$ fs was chosen in calculating charge conductivity of $\sigma$ and black dashed line indicates the experimental value. The $x$ and $y$ directions indicate $[21\bar{5}]$ and $[\bar{1}20]$, respectively.

**Table S1.** Charge conductivity and spin Hall conductivity tensor of (001)-oriented and (211)-oriented Ni$_4$W at Fermi level. The tensor of (211) film was obtained as follows $\sigma_{(211)}{}^k_{ij} = \sum_{l,m,n} D_{il} D_{jm} D_{kn} \sigma_{(001)}{}^n_{lm}$, where $D$ is a rotation matrix[8].

| | Charge conductivity at E = E$_F$ ($10^4$ S/cm), $\tau = 5$ fs | Spin Hall conductivity at E = E$_F$ (($\hbar/2e$)(S/cm)) | | |
|---|---|---|---|---|
| | $\begin{pmatrix} \sigma_{xx} & \sigma_{xy} & \sigma_{xz} \\ \sigma_{yx} & \sigma_{yy} & \sigma_{yz} \\ \sigma_{zx} & \sigma_{zy} & \sigma_{zz} \end{pmatrix}$ | $\begin{pmatrix} \sigma^x_{xx} & \sigma^x_{xy} & \sigma^x_{xz} \\ \sigma^x_{yx} & \sigma^x_{yy} & \sigma^x_{yz} \\ \sigma^x_{zx} & \sigma^x_{zy} & \sigma^x_{zz} \end{pmatrix}$ | $\begin{pmatrix} \sigma^y_{xx} & \sigma^y_{xy} & \sigma^y_{xz} \\ \sigma^y_{yx} & \sigma^y_{yy} & \sigma^y_{yz} \\ \sigma^y_{zx} & \sigma^y_{zy} & \sigma^y_{zz} \end{pmatrix}$ | $\begin{pmatrix} \sigma^z_{xx} & \sigma^z_{xy} & \sigma^z_{xz} \\ \sigma^z_{yx} & \sigma^z_{yy} & \sigma^z_{yz} \\ \sigma^z_{zx} & \sigma^z_{zy} & \sigma^z_{zz} \end{pmatrix}$ |
| Ni$_4$W(001) $x$ [100] $y$ [010] | $\begin{pmatrix} 0.71 & 0 & 0 \\ 0 & 0.71 & 0 \\ 0 & 0 & 1.32 \end{pmatrix}$ | $\begin{pmatrix} 0 & 0 & 161 \\ 0 & 0 & 234 \\ -119 & -163 & 0 \end{pmatrix}$ | $\begin{pmatrix} 0 & 0 & -234 \\ 0 & 0 & 161 \\ 163 & -119 & 0 \end{pmatrix}$ | $\begin{pmatrix} 49 & 1364 & 0 \\ -1364 & 49 & 0 \\ 0 & 0 & -23 \end{pmatrix}$ |
| Ni$_4$W(211) $x$ [21$\bar{5}$] $y$ [$\bar{1}$20] | $\begin{pmatrix} 1.12 & 0 & -0.29 \\ 0 & 0.71 & 0 \\ -0.29 & 0 & 0.92 \end{pmatrix}$ | $\begin{pmatrix} -12 & -570 & 50 \\ 536 & -40 & 978 \\ 114 & -954 & 8 \end{pmatrix}$ | $\begin{pmatrix} 34 & 96 & -188 \\ -130 & 0 & 94 \\ 210 & -70 & -34 \end{pmatrix}$ | $\begin{pmatrix} -16 & 570 & 128 \\ -620 & 28 & -536 \\ -100 & 570 & 30 \end{pmatrix}$ |

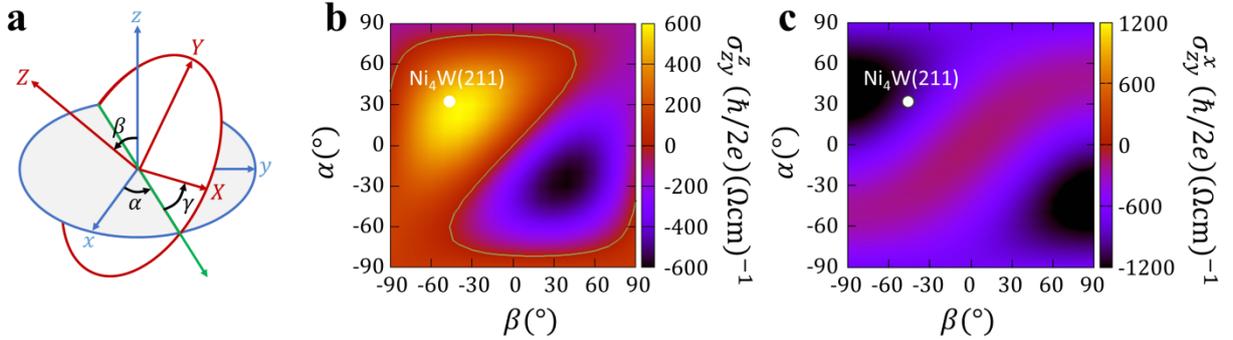

**Figure S2. a,** Schematic diagram of Euler angles employed in the current study. The $x$, $y$, and $z$ and $X$, $Y$, and $Z$ indicate coordinate axes before and after rotation, respectively. The $\alpha, \beta,$ and $\gamma$ represent three Euler angles. Here, $\alpha = 31.85°$ and $\beta = -46.54°$ were used to rotate Ni$_4$W-orientation from (001) to (211). The $\gamma$ determines the azimuthal angle of the rotated (211) orientation. The $\gamma$ was chosen to be $-40.46°$, which guarantees $X \parallel [21\bar{5}]$ and $Y \parallel [\bar{1}20]$. **b,c,** Calculated $\sigma^z_{zy}$ and $\sigma^x_{zy}$ of the rotated Ni$_4$W as a function of $\alpha$ and $\beta$ with $\gamma = -40.46°$. In both figures, green dots indicate the (211) orientation. For a given $\gamma$, both conventional and unconventional spin hall conductivity tensor components are nearly optimal at the (211) orientation, indicating that (211) direction is close to optimal in switching efficiency.

## 3. Composition determination by Rutherford backscattering spectrometry

Rutherford backscattering spectrometry (RBS) was used to evaluate the elemental composition of the Ni-W film. A National Electrostatics MAS 1700 pelletron tandem ion accelerator (5SDH) was used to generate a 4 nA beam of 2.0 MeV He+ through a 2x2 mm aperture. Data acquisition employed a Charles Evans analytical end station equipped with an Ortec ion detector (solid angle 3.6 msr) at 165 degree scattering angle. Beam current on sample was integrated to 5.2 uC total charge per spectrum using an Ortec charge integrator. A two-axis goniometer provided a 3 degree polar angle (sample tilt) and a precession of sample normal around the beam once per spectrum in 2 degree azimuthal steps, to eliminate possible ion channeling in substrate (the signal of which served as an independent check of charge integration, via spectral simulation).

The elemental composition of the Ni-W film prepared by co-sputtering is evaluated to be around 4:1 by analyzing the RBS data, as shown in Figure S3.

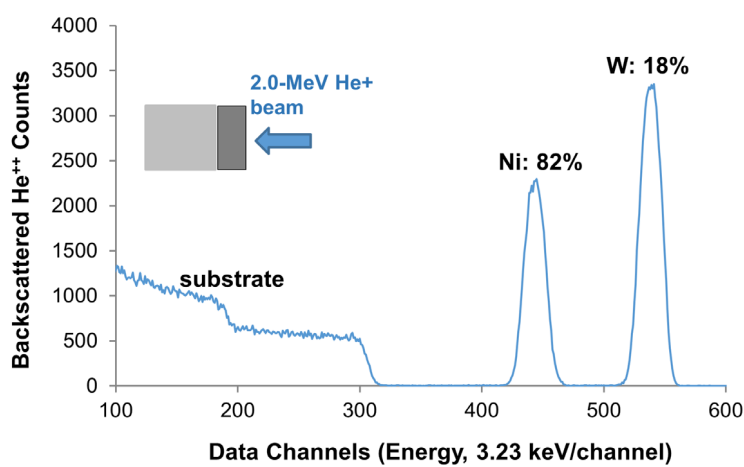

**Figure S3** RBS spectrum of a 55 nm thin film of Ni-W on $Al_2O_3(0001)$ substrate, with labeled peaks for Ni and W. The percentages of Ni and W were calculated from the relative peak integrations (each normalized to atomic number squared).

## 4. Magnetic property of $Ni_4W$ and $Ni_4W/CoFeB$

Vibrating-sample magnetometer (VSM) was used to measure the magnetic property of $Ni_4W$ film and $Ni_4W/CoFeB$ heterostructure. For the $Ni_4W$ (55 nm) sample (Figure S4 **a**), the saturation magnetization is estimated to be 5.8 emu/cc, which is almost two orders of magnitude smaller than the value for Ni[9], indicating that $Ni_4W$ is non-magnetic. VSM measurement was also performed on a $Ni_4W$ (30 nm)/CoFeB (5 nm) sample (Figure S4 **b**) with an in-plane (IP) or out-of-plane (OOP) field sweep. In-plane anisotropy can be clearly seen. The estimated saturation magnetization is 1200 emu/cc, which is similar to the values reported by previous studies[10].

To reflect these points, we have updated the Fig. SX in the Supporting Information as below:

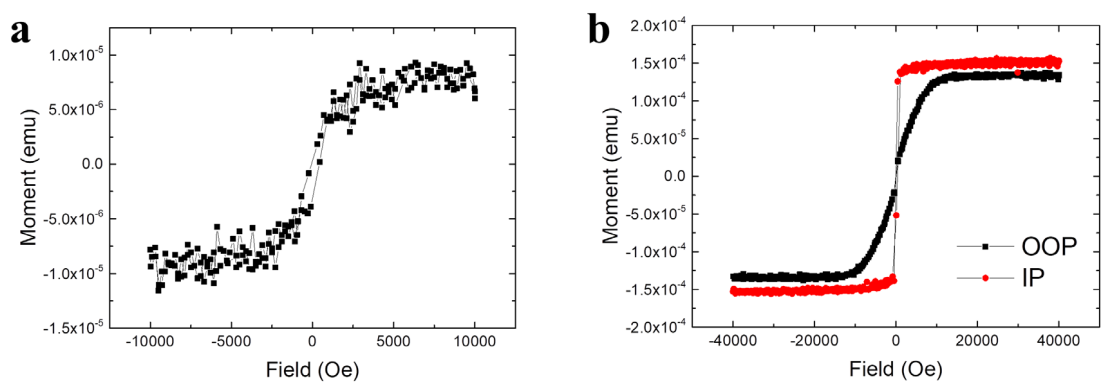

**Figure S4** Hysteresis loop of $Ni_4W$ (55nm) sample with an in-plane magnetic field (**a**) and $Ni_4W$ (30 nm)/CoFeB (5 nm) sample with in-plane or out-of-plane magnetic field (**b**).

## 5. Roughness of $Ni_4W$ surface

We used atomic force microscope (AFM) to measure the roughness of $Ni_4W$ surfaces. Figure S5 **a** shows the AFM images on a $Ni_4W$ (11 nm) sample. The root-mean-square (RMS) roughness on the area of 20 μm × 20 μm is 0.26 nm, indicating a relatively flat $Ni_4W$ surface suitable for $Ni_4W$/CoFeB heterostructure. The RMS roughness of a thicker $Ni_4W$ (55 nm) surface is 0.65 nm, as shown in Figure S5 **b**.

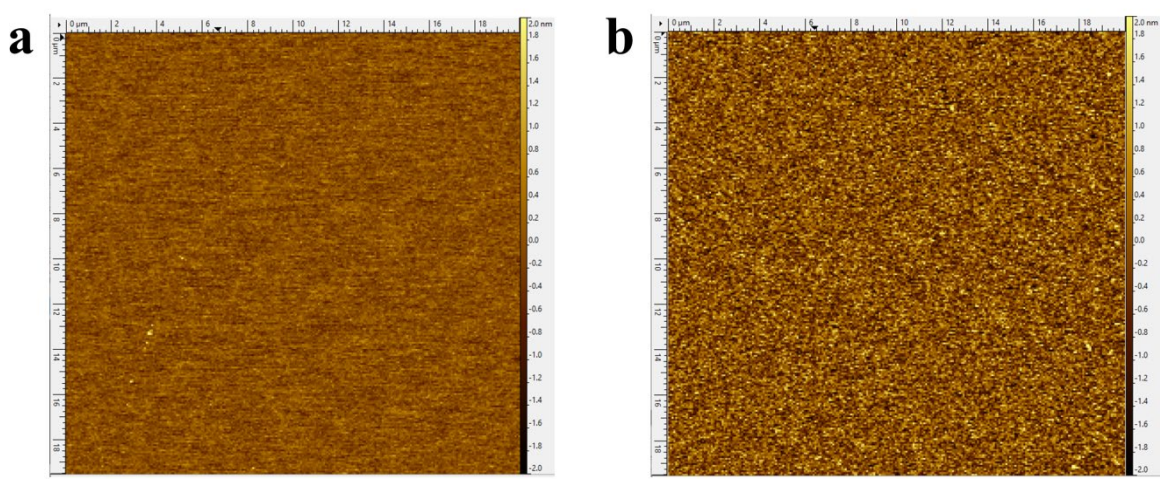

**Figure S5** AFM images on $Ni_4W$ (11 nm) (**a**) and $Ni_4W$ (55 nm) (**b**) surface.

## 6. Rocking curves of $Ni_4W$ with different thicknesses

Rocking curves of the $Ni_4W$ (211) peak for $Ni_4W$ samples with different $Ni_4W$ thicknesses are shown in Figure S6. The curves were fitted as a superposition of a broad peak and a sharp peak. By comparing the three figures in Figure S6, it is clear to see that the broad peak component increases as the Ni$_4$W thickness increases. This indicates there are more tilted and relaxed grains as the thickness of Ni$_4$W increases, which follows a similar trend as for AlN grown on sapphire substrate[11]. The full width at half maximum (FWHM) values of the sharp peaks are 0.096°, 0.103°, and 0.161° for the 11 nm, 28 nm, and 55 nm samples, respectively.

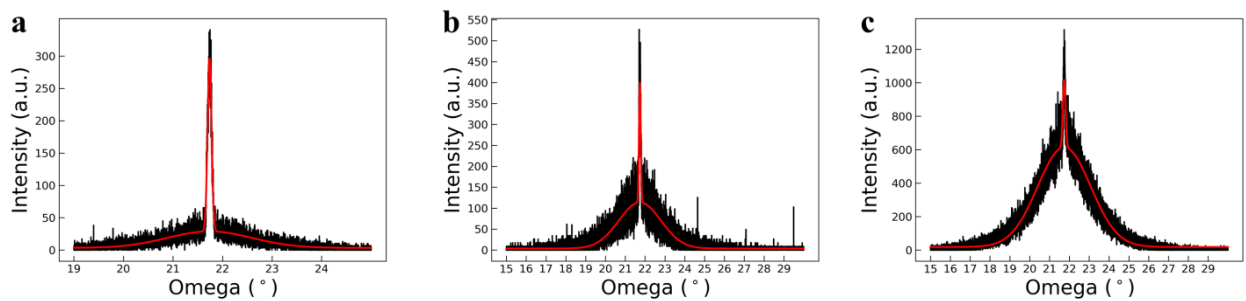

**Figure S6** Rocking curve of Ni$_4$W(211) peak with Ni$_4$W thickness of 11 nm (**a**), 28 nm (**b**) and 55 nm (**c**).

## 7. Reciprocal space mappings for W/Ni$_4$W/CFB sample

Figure S7 shows the reciprocal space mapping of Al$_2$O$_3$ (0001)/W (2 nm)/Ni$_4$W (30 nm)/CoFeB (5 nm)/cap with different axes combinations from the main text. These mappings collectively confirm the sharply textured features of both W and Ni$_4$W layers for all orientations.

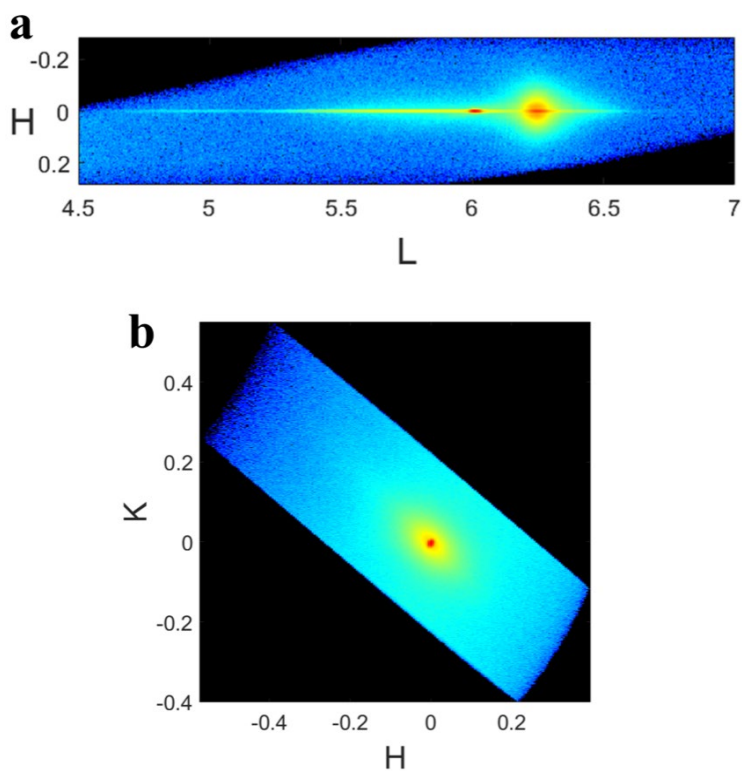

**Figure S7** Reciprocal space mapping of Al$_2$O$_3$ (0001)/W (2 nm)/Ni$_4$W (30 nm)/CoFeB (5 nm)/cap in units of sapphire plotted in axes of H and L (**a**), H and K (**b**).

## 8. TEM characterization for W/Ni$_4$W/CFB sample

A low-magnification high-angle annular-dark-field (HAADF) scanning transmission electron microscopy (STEM) image of the Al$_2$O$_3$(0001)/W (2 nm)/Ni$_4$W (30 nm)/CoFeB (5 nm)/Ta cross-section sample at the Al$_2$O$_3$ [10$\bar{1}$0] zone axis orientation is shown in the left subfigure of Figure S8 **a** (beam energy 200 keV, beam current 320 pA, convergence semi-angle 10 mrad, detector semi-angle range 59-200 mrad). Clear differences in atomic number contrast can be seen across each of the labeled layers.

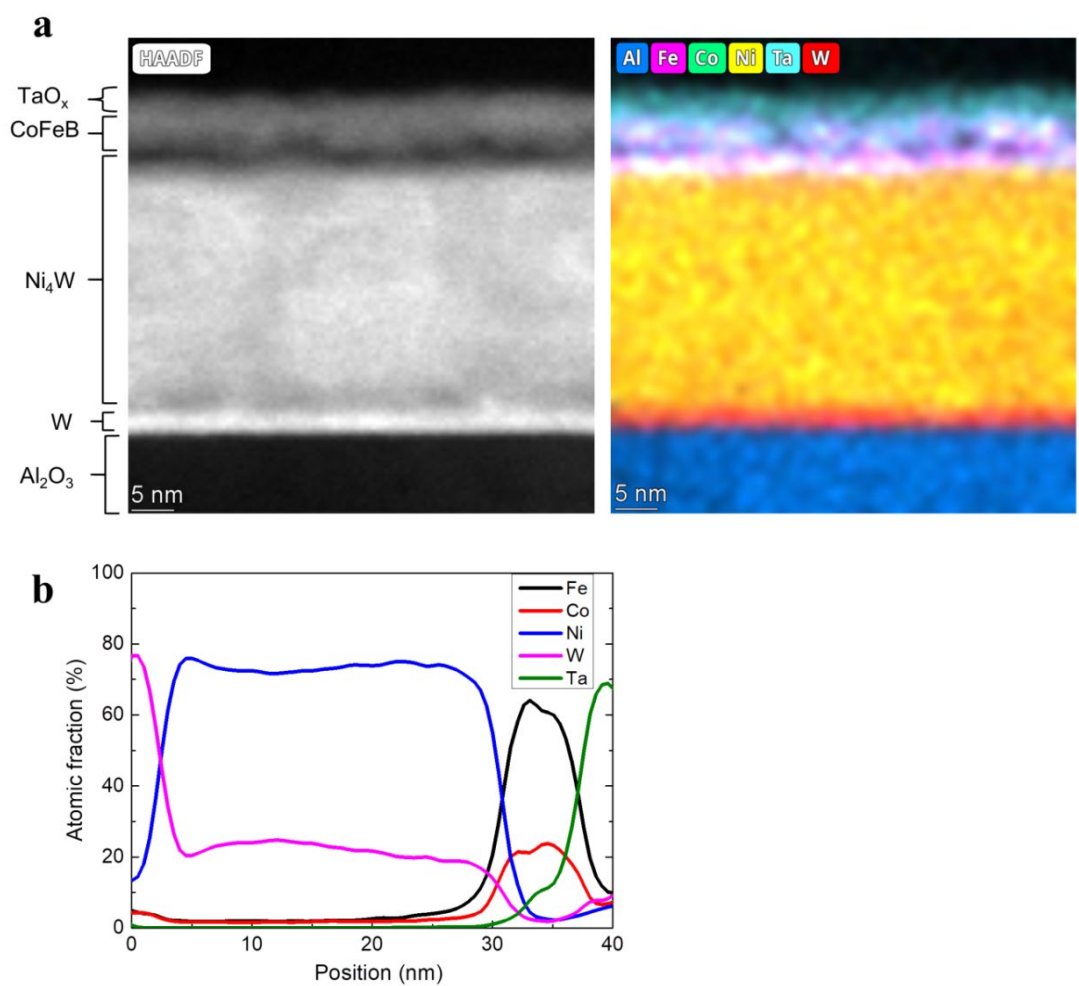

**Figure S8 a**, Low-magnification HAADF-STEM image and STEM-EDS element maps of Al$_2$O$_3$(0001)/W (3 nm)/Ni$_4$W (30 nm)/CoFeB (5 nm)/Ta sample. **b**, Elemental line profile across the stack processed from the STEM-EDS area map.

In parallel with HAADF imaging, energy-dispersive X-ray spectroscopy (EDS) mapping, shown in the right subfigure of Figure S8 **a**, was used to produce an elemental map of the same region (note that a raw pixel size of 0.5 nm blurred by a 1.5 x 1.5 nm square average filter to improve fitting of the net characteristic peak counts in the EDS spectra). Employing standardless Brown-Powell quantification[12] without absorption corrections, the EDS map was binned into the atomic percentage linescan profile of Figure S8 **b**. Throughout the $Ni_4W$ film, the measured atomic ratio of Ni and W is close to 4:1 (within the uncertainty of standardless quantification applied to a FIB liftout sample at a high-symmetry zone axis orientation). Measured layer thicknesses closely match their nominal expected values. Due to effects of STEM beam broadening, the extent of inter-diffusion at the $W/Ni_4W$ and $Ni_4W/CFB$ interfaces is unknown, but there is no evidence of strong intermixing between any of the layers.

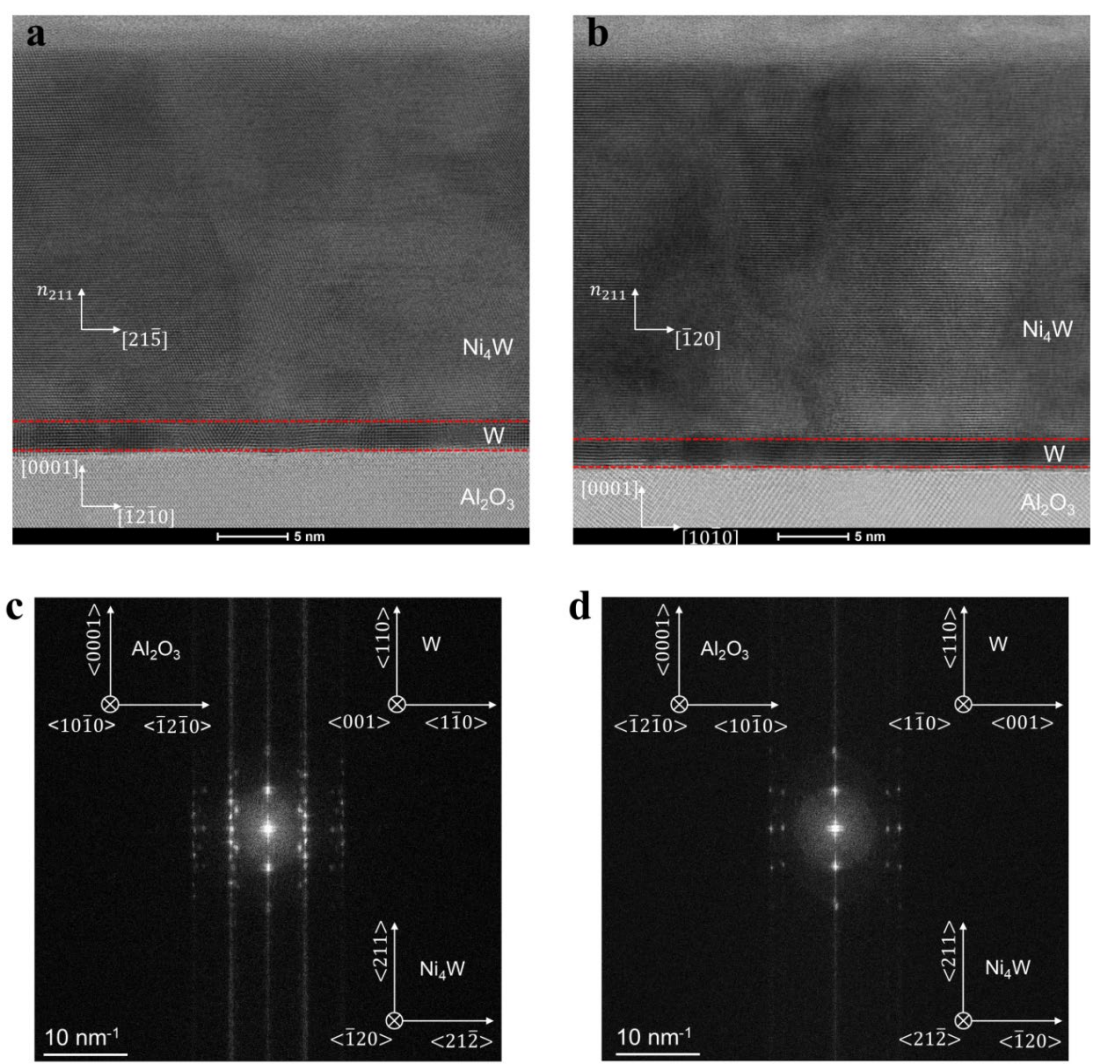

**Figure S9 a**, **b**, Bright-field HRSTEM images of $Al_2O_3(0001)/W$ (2 nm)/$Ni_4W$ (30 nm)/CoFeB (5 nm)/Ta sample oriented to zone axes $Al_2O_3$ $[10\bar{1}0]$ (**a**) and $Al_2O_3$ $[\bar{1}2\bar{1}0]$ (**b**). The observed predominant texture axes of $Ni_4W$ and $Al_2O_3$ are shown. Red dashed lines are used for layer separation. $n_{211}$ is the film normal vector for the $Ni_4W$ (211) planes. **c**, **d**, FFT obtained from **a**, **b**, respectively. Note that whereas **a,b** use the real-space vectors as throughout the paper, in **c,d**, these are reciprocal space vectors that are perpendicular to corresponding planes.

Aberration-corrected high-resolution STEM (HRSTEM) imaging was used to probe the atomic structure of the film stack and determine the orientation relationships between $Al_2O_3$, W and $Ni_4W$. The lattice structure of the film stack is imaged for two different cross-section zone axes using bright-field STEM imaging in Figure S9 **a**, **b** (beam energy 300 keV, beam current 50 pA, convergence semi-angle 21 mrad, detector semi-angle 13 mrad). Notwithstanding moderate surface damage from the FIB liftout and ambient oxidation, the $Ni_4W$ layer exhibits extremely high quality: sharp in-plane and out-of-plane texture, large grain sizes, and highly coherent grain boundaries ($Ni_4W$ grain morphology and in-plane texture can also be seen in a plan-view image, see Figure S10 **b**). The W layer is also of high quality by virtue of having sharp texture, however it exhibits extensive grain rotation with incoherent grain boundaries, most clearly visible in Figure S9 **a**. (W seed layer grain morphology and in-plane texture can be seen more directly in a plan-view image, see Figure S10 **a**). Fast-Fourier transformation (FFT) of HRSTEM images is shown Figure S9 **c**, **d**. The diffraction-like dot pattern of the FFT confirms the high crystalline quality and sharp texture of the films. Relative to Figure S9 **d**, there is a more intricate set of dots in Figure S9 **c**; this is attributable to clearer imaging of the higher-symmetry W grain orientations as well as scattered mirror twinning in the $Ni_4W$ layer, traits that are also visible in the real space STEM image (Figure S9 **a**).

Texture/orientation relationships and material phases were confirmed from HRSTEM data correlated with selected-area electron diffraction (SAED) data (SAED not presented here). As labeled throughout Figure S9, the $Ni_4W$ exhibits (211) orientation, while W exhibits (110) orientation along the out-of-plane $Al_2O_3$ <0001> direction. For in-plane texture, (211) $Ni_4W$ is elegantly quasi-hexagonal: its <$\bar{1}20$> direction – normal to ($\bar{1}20$) plane – aligns with $Al_2O_3$ <$10\bar{1}0$>, while its <$21\bar{5}$> direction – corresponding to ($21\bar{2}$) plane – aligns with $Al_2O_3$ <$\bar{1}2\bar{1}0$> direction. The $Ni_4W$ texture is transferred from the substrate via rotations of rectangular-symmetry (110) W grains: in the predominant texture, the W <100> directions align with $Al_2O_3$ <$10\bar{1}0$> and their <$1\bar{1}0$> directions align with $Al_2O_3$ <$\bar{1}2\bar{1}0$>. As summarized in Table S2, when analyzing FFT dot patterns of HRSTEM images, the measured d-spacing values of various planes of sapphire $Al_2O_3$, body-centered-cubic W, and body-centered tetragonal $Ni_4W$ all agree with reference crystallographic data within expected 3% error (error arises from limited calibration precision and hysteresis; SAED data, obtained for these and other Bragg reflections, matches equally well).

**Table S2.** Measured d-spacing of various planes of Al$_2$O$_3$, W and Ni$_4$W extracted from FFT patterns of HRSTEM images at the Al$_2$O$_3$ [$\bar{1}2\bar{1}0$] zone axis orientation of Al$_2$O$_3$(0001)/W (2 nm)/Ni$_4$W (30 nm)/CoFeB (5 nm)/Ta, including comparison to reference crystallographic data.

| Lattice plane | TEM spatial frequency (nm$^{-1}$) | TEM d-spacing (nm) | d-spacing from database (nm) | d-spacing difference (%) |
|---|---|---|---|---|
| Al$_2$O$_3$ {0006} | 0.45 | 0.222 | 0.217 | 2.30 |
| Al$_2$O$_3$ {0-330} | 0.71 | 0.141 | 0.137 | 2.92 |
| W {110} | 0.44 | 0.227 | 0.224 | 1.34 |
| W {002} | 0.64 | 0.156 | 0.158 | -1.27 |
| W {211} | 0.77 | 0.130 | 0.129 | 0.78 |
| Ni$_4$W {211} | 0.48 | 0.208 | 0.208 | 0 |
| Ni$_4$W {240} | 0.78 | 0.128 | 0.128 | 0 |

To more fully study the morphology of W and Ni$_4$W layers, we prepared a plan-view TEM specimen (FIB liftout rotated and then thinned such that the incident TEM beam can be parallel to the film normal[13]) from the Al$_2$O$_3$(0001)/W (2 nm)/Ni$_4$W (30 nm)/CoFeB (5 nm)/Ta sample. The end result has the electron beam viewing down the Al$_2$O$_3$ [0001] zone axis, bottom-up from substrate through the whole film stack. By examining regions milled to different depths in the FIB plan-view sample, HAADF-STEM imaging (beam energy 300 keV, beam current 50 pA, convergence semi-angle 21 mrad, detector semi-angle range 58-200 mrad) reveals in-plane details of W and Ni$_4$W layers, as shown in Figure S10 **a**, **b**, respectively. The W seed layer is comprised of polydisperse grains with incoherent boundaries, however the grains predominantly follow the Al$_2$O$_3$ substrate and produce six-fold symmetry in the FFT, as seen in Figure S10 **a**. The Ni$_4$W film accommodates all three rotations of the W seed grain equivalently well and exhibits highly coherent grain boundaries, resulting in large domains with very slight rotational variation; this unidirectional seamless in-plane structure of Ni$_4$W is seen in the image and associated FFT dot profile of Figure S10 **b**.

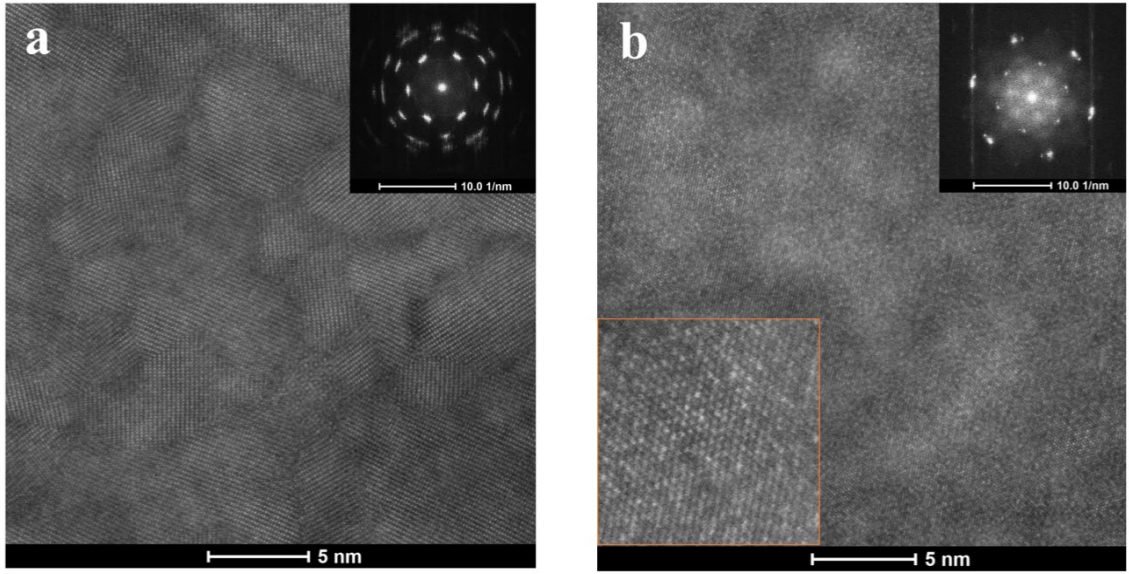

**Figure S10 a**, **b**, HAADF HRSTEM images of Al$_2$O$_3$(0001)/W (2 nm)/Ni$_4$W (30 nm)/CoFeB (5 nm)/Ta oriented to the Al$_2$O$_3$ [0001] zone axis, from regions respectively exposing W (**a**) and Ni$_4$W (**b**). The top-right inset of each image is the corresponding FFT. A zoomed-in view of Ni$_4$W serves as the bottom-left inset of that image.

## 9. Second harmonic Hall analysis

Second harmonic Hall measurements were used to evaluate the spin-orbit torque (SOT) efficiency of Ni$_4$W. An a.c. current with amplitude of 15 mA and frequency of 133.7 Hz is applied in the longitudinal direction of the Hall bar (Figure S11 **a**) Hall bar devices are patterned with the current channel aligned at various angles relative to the Ni$_4$W <21$\bar{5}$> axis, which is denoted as ψ. An external magnetic field between 3,000 Oe and 1 Tesla is applied at an angle of φ relative to the current direction. The first and second harmonic Hall signals are measured by an lock-in amplifier (LIA).

The first harmonic Hall voltage can be has dependence on the magnetic field angle as[8, 14]

$$V_\omega = V_P \sin 2\varphi \qquad (1)$$

when the magnetic field is applied in the plane of the sample. $V_P$ is the planar Hall effect (PHE) voltage. The second harmonic Hall voltage can be expressed as

$$V_{2\omega} = V_{DL}^Y \cos\varphi + V_{DL}^X \sin\varphi + V_{DL}^Z \cos 2\varphi + V_{FL}^Y \cos\varphi \cos 2\varphi + V_{FL}^X \sin\varphi \cos 2\varphi + V_{PNE} \sin 2\varphi + V_{FL}^Z \qquad (2)$$

where $V_{DL}^X, V_{DL}^Y, V_{DL}^Z$ are voltage components from damping-like torques generated by X, Y and Z spins, respectively, and $V_{FL}^X, V_{FL}^Y, V_{FL}^Z$ are the field-like counterparts. $V_{PNE}$ is the planar Nernst effect (PNE) voltage.

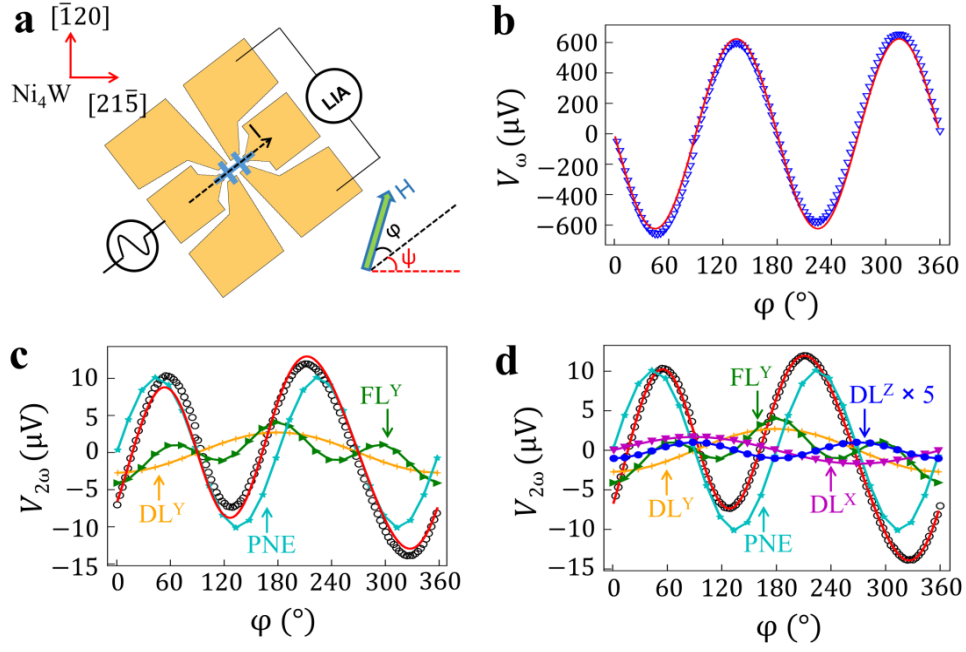

**Figure S11 a**, Schematic of the Hall bar device for second harmonic Hall measurements. An external field is applied with an angle of φ with respect to the current flow direction. The current flow direction has an angle of ψ relative to $Ni_4W$ <$21\bar{5}$> axis. **b-d**, First harmonic Hall signal (**b**) and second harmonic Hall signals (**c,d**) of $Ni_4W$ (5 nm)/CoFeB (5 nm) under a magnetic field of 4,000 Oe with current along $Ni_4W$ <$\bar{1}20$> (ψ = 90°). Second harmonic Hall signal is fitted with only Y-spin component (**c**) and with X-, Y-, Z-spin components (**d**).

Figure S11 **b** shows the first harmonic Hall signal of $Ni_4W$ (5 nm)/CoFeB (5 nm), with current flowing along $Ni_4W$ <$\bar{1}20$> direction (ψ = 90°) and magnetic field of 4,000 Oe. To fit the second harmonic Hall data, we first use only Y-spin components ($V_{DL}^Y$ and $V_{FL}^Y$) and PNE ($V_{PNE}$), as shown in Figure S11 **c**. It is clear that only considering Y-spin cannot fit the data well. We therefore include the X- and Z-spin components for the fitting, the result of which is shown in Figure S11 **d**. The non-zero $V_{DL}^X$ and $V_{DL}^Z$ confirm the existence of X- and Z-spin, respectively.

To evaluate the SOT efficiencies of X-, Y- and Z-spins, we carried out the field dependence study of the harmonic Hall measurements. The voltage components are linearly dependent on the inverse of magnetic field[8, 10b, 14]:

$$V_{DL}^X = -V_A \frac{H_{DL}^X}{2(H+H_K)} \qquad (3)$$

$$V_{FL}^X = V_P \frac{H_{FL}^X}{H} \qquad (4)$$

$$V_{DL}^Y = -V_A \frac{H_{DL}^Y}{2(H+H_K)} + V_{ANE} \qquad (5)$$

$$V_{FL}^Y = V_P \frac{H_{FL}^Y}{H} \qquad (6)$$

$$V_{DL}^Z = V_P \frac{H_{DL}^Z}{H} \qquad (7)$$

$$V_{FL}^Z = -V_A \frac{H_{FL}^Z}{2(H+H_K)} + C \qquad (8)$$

$H_{DL}^X, H_{DL}^Y, H_{DL}^Z, H_{FL}^X, H_{FL}^Y, H_{FL}^Z$ are effective fields associated with damping-like and field-like torques from X-, Y-, Z-spins, respectively. $V_A$ is the anomalous Hall voltage and $V_P$ is planar Hall voltage. $V_{ANE}$ is anomalous Nernst voltage. $H$ is the external field and $H_K$ is anisotropic field. $C$ is a constant.

SOT efficiencies are calculated by

$$\theta_{DL(FL)}^i = \frac{2e}{\hbar} M_S t_{FM} \frac{\mu_0 H_{DL(FL)}^i}{J_{NM}} \qquad (9)$$

where e is electron charge, $\hbar$ is reduced Planck constant, $M_S$ is saturation magnetization, $t_{FM}$ is the thickness of ferromagnetic (FM) layer, and $\mu_0$ is vacuum magnetic permeability. $H_{DL(FL)}^i$ is the current-induced damping- or field-like effective field from i-spin (i=X,Y,Z), and $J_{NM}$ is the charge current density in the non-magnetic (NM) layer. The Oersted field generated by current in NM layer is removed from $H_{FL}^Y$. The value of Oersted field is estimated by $H_{Oe} = \frac{1}{2} J_{NM} t_{NM}$[8], where $t_{NM}$ is the thickness of NM layer.

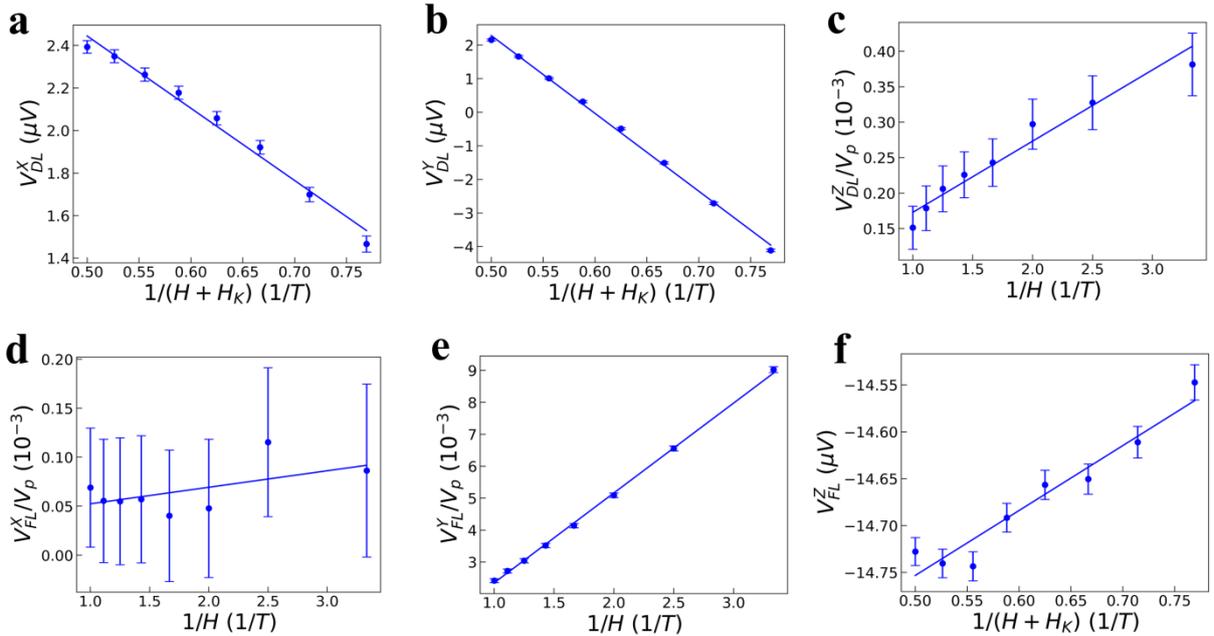

**Figure S12**. Field dependence of the second harmonic Hall voltage components of Ni$_4$W (5 nm)/CoFeB (5 nm) with current along Ni$_4$W <$\bar{1}$20> (ψ = 90°). Linear trend can be seen for all components except $V_{FL}^X$, which indicates the magnitude is negligible.

As shown in Figure S12, all the voltage components except $V_{FL}^X$ show a linear dependence on the inverse of field, indicating a non-zero damping- or field-like SOT from the corresponding spin. In contrast, unconventional voltage components ($V_{DL}^X, V_{DL}^Z, V_{FL}^X, V_{FL}^Z$) are negligible in the reference sample W (3 nm)/CoFeB (5 nm), as shown in Figure S13 **b**. $\theta_{DL}^Y$ and $\theta_{FL}^Y$ is calculated to be $-0.132 \pm 0.003$ and $0.036 \pm 0.001$ using the field dependence (Figure S13 **c,d**), similar to the values reported[15].

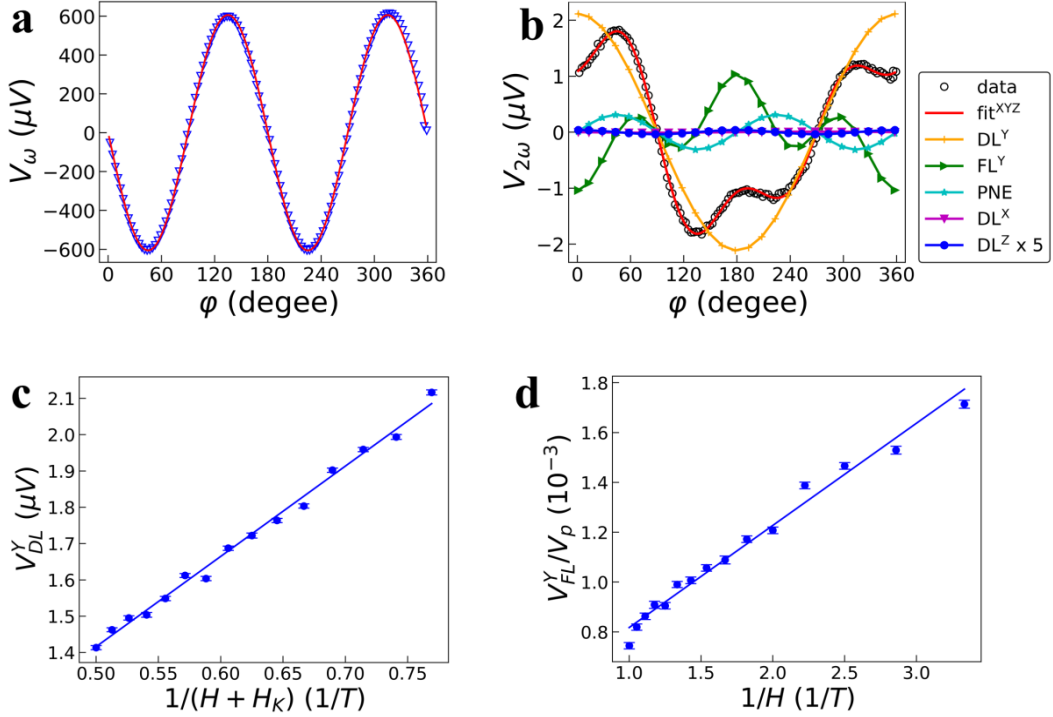

**Figure S13 a,b**, First harmonic Hall signal (**a**) and second harmonic Hall signals (**b**) of $Al_2O_3$/W (3 nm)/CoFeB (5 nm)/cap under a magnetic field of 3,000 Oe. Second harmonic Hall signal is fitted with X-, Y-, Z-spin components, but no unconventional components can be observed. **c,d**, field dependence of second harmonic Hall voltage components $V_{DL}^Y$ (**c**) and $V_{FL}^Y$ (**d**).

We performed harmonic Hall measurements for different current directions, as shown in Figure S14. The damping-like SOT efficiencies show clear dependence on the current direction ψ (see Figure S11 **a** for definition). As discussed in the main text, the large $\theta_{DL}^X$ and $\theta_{DL}^Y$ in $Ni_4W$ (5 nm)/CoFeB (5 nm) can possibly originate from Rashba effects. We note that in addition to the intrinsic SHC from $Ni_4W$ that has strong dependence on current direction[8,16], Rashba effects can also exhibit anisotropic behavior as discovered in single crystals[17].

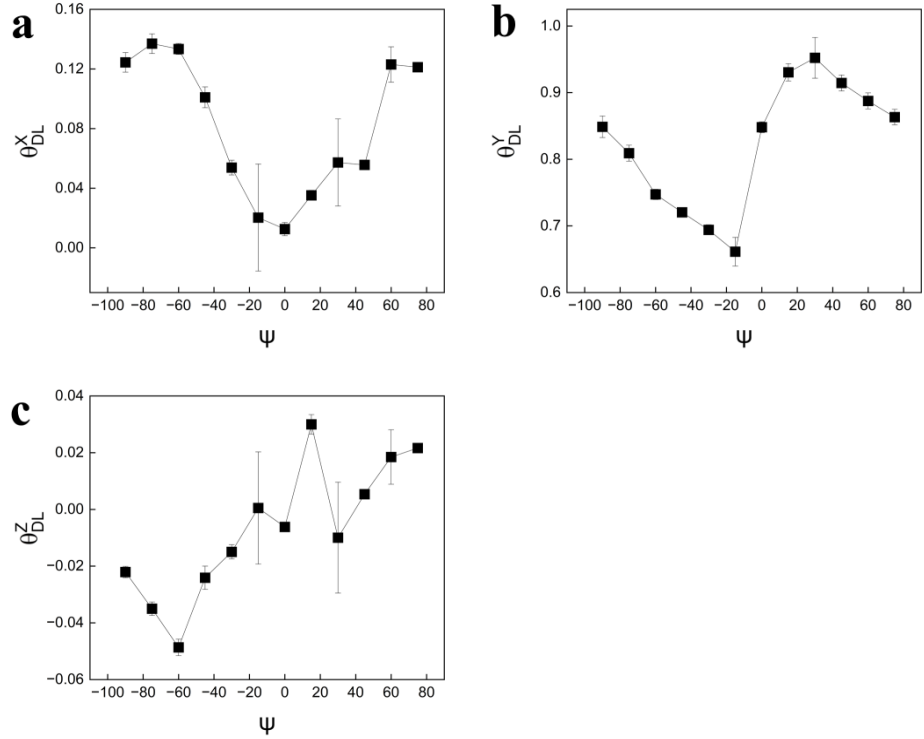

**Figure S14** Damping-like SOT efficiencies as a function of current direction of Ni$_4$W (5 nm)/CoFeB (5 nm) ($\psi$ is the angle relative to Ni$_4$W <21$\bar{5}$> axis).

## 10. Macro-spin simulations

**Table S3.** Summary of SOT efficiencies of Pt, W and recently discovered unconventional SOT materials.

| Materials | $\theta_{DL}^X/\theta_{FL}^X$ | $\theta_{DL}^Y/\theta_{FL}^Y$ | $\theta_{DL}^Z/\theta_{FL}^Z$ | $\sigma_{DL}^Z$ ($\hbar/2e$) ($\Omega$m)$^{-1}$ | $\rho$($\mu\Omega$ cm) |
|---|---|---|---|---|---|
| Pt [18] | - / - | 0.10 / -0.004 | - / - | 0 | 20 |
| W | - / - | -0.286 / 0.013 | - / - | 0 | 117 |
| WTe$_2$ [19] | - / - | 0.030 / 0.034 | 0.013 / - | 3.6*10$^3$ | 380 |
| Mn$_3$GaN [20] | -0.013 / - | 0.025 / - | 0.019 / -0.15 | 8.6*10$^3$ | 220 |
| MnPd$_3$ [14c] | 0.007 / - | 0.30 / 0.03 | 0.009 / - | 1.42*10$^4$ | 60 |
| RuO$_2$ [8] | - / 0.002 | 0.05 / 0.004 | 0.010 / - | 7.0*10$^3$ | 140 |
| TaIrTe$_4$ [21] | - / - | 0.113 / - | 0.043 / - | 2.07*10$^4$ | 209.7 |
| PtTe$_2$/WTe$_2$ [22] | - / - | 0.147 / - | 0.013 / - | 2.0*10$^4$ | 62.5 |
| **Ni$_4$W** | **0.124 / -** | **0.849 / 0.577** | **-0.022 / 0.025** | **1.47*10$^4$** | **150** |

Table S3 summarizes the SOT efficiencies of Pt[18], W and state-of-the-art unconventional SOT materials[8, 14c, 19-20],[21-22]. Ni$_4$W (5 nm) exhibits an unconventional damping-like SHC value $\sigma_{DL}^Z$ that is comparable to these materials. Additionally, its conventional damping-like efficiency ($\theta_{DL}^Y$) is much larger than those of previously reported materials. Its sizable field-like SOT efficiency ($\theta_{FL}^Y$) can also potentially assist the switching of perpendicular magnetization[23]. The resistivity of Ni$_4$W is 150 μΩ cm, which can effectively avoid the shunting effect for SOT channel/ferromagnet bilayer structure in memory or logic devices.

The Landau-Lifshitz-Gilbert (LLG) equation governing the magnetization dynamics is

$$\frac{d\mathbf{m}}{dt} = -\gamma \mathbf{m} \times \mathbf{H}_{\text{eff}} + \alpha \mathbf{m} \times \frac{d\mathbf{m}}{dt} + \frac{\gamma}{\mu_0 M_S d} \mathbf{N}$$

where $\mathbf{m}$ is the unit vector along the magnetization direction, with saturation value $M_S = 1.2 \times 10^6$ A/m, $\gamma$ is the gyromagnetic ratio, $\alpha = 0.001$ is the intrinsic damping parameter, $\mu_0$ is the vacuum permeability and $d = 5$ nm is the thickness of the ferromagnetic thin film. The perpendicular magnetic anisotropy is described by means of the effective field, $\mathbf{H}_{\text{eff}} = H_{\text{eff}}(\mathbf{m} \cdot \mathbf{Z})\mathbf{Z}$, where $\mathbf{Z}$ is the out-of-plane direction. The effective anisotropy field magnitude is assumed to be $H_{\text{eff}} = 7957 \times 10^3$ A/m. The total spin-orbit toque, $\mathbf{N}$, comprises field-like and damping-like torque due to all spin components:

$$\mathbf{N} = \sum_{j=X,Y,Z} \tau_{FL}^j \mathbf{m} \times \mathbf{e}_j + \tau_{DL}^j \mathbf{m} \times (\mathbf{m} \times \mathbf{e}_j),$$

where $\mathbf{e}_{X,Y,Z} = \mathbf{X}, \mathbf{Y}, \mathbf{Z}$, refer to the unit vectors along the three mutually orthogonal components. The magnitudes of damping-like and field like torques are taken to be $\tau_{DL(FL)}^j = (\hbar/2e)\theta_{DL(FL)}^j J$, where the Hall angles are provided in Table S3 and the charge current density is assumed to be $J = 3 \times 10^7$ A/cm² for the comparison of the switching efficiency calculations among all unconventional SOT materials. $J = 9 \times 10^6$ A/cm² is used for the calculations where we considered the impact of distinct SOT components of Ni$_4$W, in the main text.

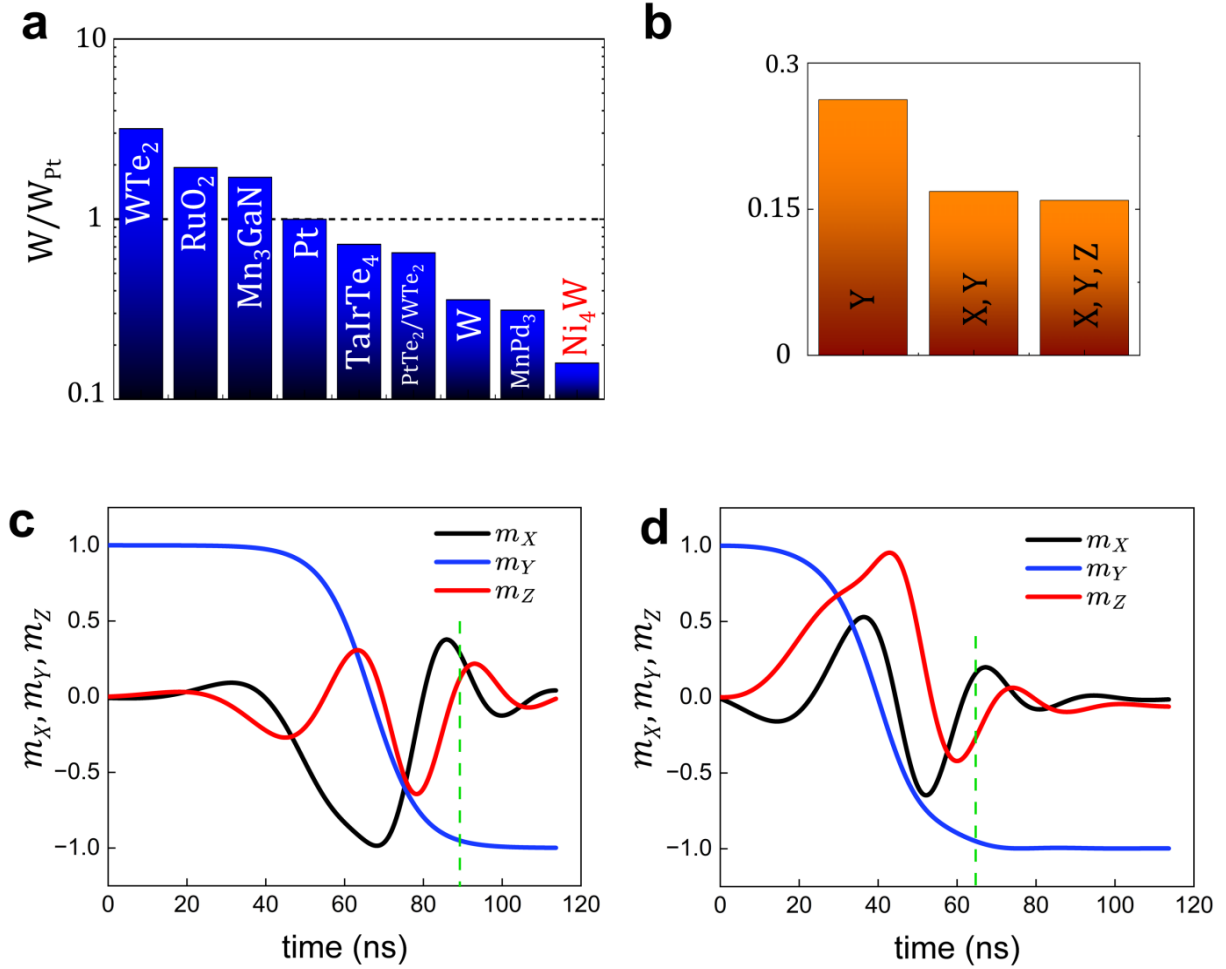

**Figure S15** Macro-spin simulations with magnetic anisotropy along $\mathbf{e}_y$. **a**, Energy delay products of SOT materials in Table S3 normalized by Pt value. **b**, Energy delay product of Ni$_4$W with different combinations of spin components. **c,d**, Magnetic dynamics using Ni$_4$W as SOT material with Y-spin only (**c**) and all X-, Y-, Z-spins (**d**).

We also performed macro-spin simulations assuming an in-plane magnetic anisotropy along $\mathbf{e}_Y$. The energy delay products of materials summarized in Table S3 are compared in Figure S15. The switching time $\tau_{sw}$ is obtained numerically by registering the time spent for full magnetization reversal according to the criteria $m_Y(\tau_{sw}) < -0.1$ and $(dm_Y/dt)_{t=\tau_{sw}} <$ 0.01, where $m_Y$ refers to the $\mathbf{e}_Y$ component of the magnetization that corresponds to the direction of the magnetic anisotropy, and the injected current density is taken to be $J = 10^6$ Acm$^{-2}$. We find that other spin components also provide efficiency enhancement in the case of Ni$_4$W. This is shown in Figure S15, where we analyze the impact of having distinct unconventional SOTs due other spin components on top of the conventional Y-spin. Here, the relative SOT switching efficiency solely through the Y-spin of Ni$_4$W is comparable to that of MnPd$_3$, but enhances as the other spin components are added. Figure S15 **c** and **d** illustrate the magnetic dynamics for Ni$_4$W assuming Y-spin only or all three spin components, respectively. The $\tau_{sw}$ is reduced from 90 ns to 65 ns when X- and Z-spins are added to the conventional Y-spin. This indicates the unconventional spin components can improve the switching efficiency for in-plane magnets.

**Supplementary References**